\documentclass[10pt,journal,compsoc]{IEEEtran}
\usepackage{graphicx}
\usepackage{multirow}
\usepackage[table,xcdraw,dvipsnames]{xcolor}
\usepackage{longtable}
\usepackage{color}
\usepackage{xcolor,colortbl}
\usepackage{tikz}

\usepackage{pifont}
\usepackage{comment}
\usepackage{ragged2e}
\usepackage{cite}
\usepackage{booktabs}
\usepackage{multirow}
\usepackage{amssymb}
\usepackage{tabularx}
\usepackage[numbers,sort&compress]{natbib}

\usepackage[dvipsnames]{xcolor}
\usepackage{soul}

\usepackage[activate]{microtype}
\usepackage[breaklinks=true,colorlinks,bookmarks=false,citecolor=blue,linkcolor=blue]{hyperref}
\usepackage[activate]{microtype}
\usepackage{svg}
\usepackage{xcolor}
\definecolor{darkgreen}{RGB}{0,150,0}
\usepackage{pifont}
\newcommand{\cmark}{\textcolor{darkgreen}{\ding{51}}}
\newcommand{\xmark}{\textcolor{red}{\ding{55}}}

\definecolor{greencheck}{RGB}{0,128,0}
\definecolor{redcross}{RGB}{255,0,0}

\usepackage{authblk}
\usepackage{booktabs}

\usepackage{url}
\usepackage[dvipsnames]{xcolor}
\usepackage{soul}

\usepackage[multi-part-units = repeat]{siunitx}
\usepackage{array,amsfonts,amsmath}
\usepackage{caption}
\usepackage{verbatim}

\begin{document}

\title{Sparks of Large Audio Models: \\A Survey and Outlook}




\author{\textsf{ Siddique Latif,~%
     Moazzam Shoukat,
    Fahad Shamshad,~
    Muhammad Usama,
    Yi Ren,
    ~Heriberto Cuay\'ahuitl, 
    Wenwu Wang, Xulong Zhang,
Roberto Togneri, Erik Cambria, and Björn W. Schuller}
\thanks{Siddique Latif is with Queensland University of Technology (QUT), Australia.
E-mail: siddique.latif@usq.edu.au} 
\thanks{Moazzam Shoukat is with Emulation AI, Australia.}
\thanks{Fahad Shamshad is with Mohamed bin Zayed University of Artificial Intelligence, Abu Dhabi, UAE.}
\thanks{Muhammad Usama is with NUCES, Pakistan.}
\thanks{Yi Ren is with the Speech and Audio Team, Bytedance AI Lab, Singapore.}
\thanks{Heriberto Cuay\'ahuitl is with the University of Lincoln, UK.}
\thanks{Wenwu Wang is with the University of Surrey, UK.}
\thanks{Xulong Zhang is with Lab of Large Audio Models, Ping An Technology, China.}
\thanks{Roberto Togneri is with the University of Western Australia, Australia.}
\thanks{Erik Cambria is with Nanyang Technological University, Singapore.}
\thanks{Björn W.\ Schuller is with GLAM -- the Group on Language, Audio, $\&$ Music, Imperial College London, UK and is also with the Chair EIHW, University of Augsburg, Germany.}
%
}


\IEEEtitleabstractindextext{
\begin{abstract}

This survey paper provides a comprehensive overview of the recent advancements and challenges in applying large language models to the field of audio signal processing. Audio processing, with its diverse signal representations and a wide range of sources –- from human voices to musical instruments and environmental sounds –- poses challenges distinct from those found in traditional Natural Language Processing scenarios. Nevertheless, \textit{Large Audio Models}, epitomised by transformer-based architectures, have shown marked efficacy in this sphere. By leveraging massive amount of data, these models have demonstrated prowess in a variety of audio tasks, spanning from Automatic Speech Recognition and Text-To-Speech to Music Generation, among others. Notably, recently these Foundational Audio Models, like SeamlessM4T, have started showing abilities to act as universal translators, supporting multiple speech tasks for up to 100 languages without any reliance on separate task-specific systems. This paper presents an in-depth analysis of state-of-the-art methodologies regarding \textit{Foundational Large Audio Models}, their performance benchmarks, and their applicability to real-world scenarios. We also highlight current limitations and provide insights into potential future research directions in the realm of \textit{Large Audio Models} with the intent to spark further discussion, thereby fostering innovation in the next generation of audio-processing systems. Furthermore, to cope with the rapid development in this area, we will consistently update the relevant repository with relevant recent articles and their open-source implementations at \textbf{{\fontfamily{qcr}\selectfont{\color{magenta}\href{https://github.com/EmulationAI/awesome-large-audio-models}{\nolinkurl{https://github.com/EmulationAI/awesome-large-audio-models}}}}}.
\end{abstract}
\begin{IEEEkeywords}
Large language models, foundation models, large audio models, audio processing, speech processing, music signal processing, multimodality
\end{IEEEkeywords}}
\maketitle

\IEEEdisplaynontitleabstractindextext

\IEEEpeerreviewmaketitle

 

\IEEEraisesectionheading{\section{Introduction}\label{sec:introduction}}
\IEEEPARstart{A}{udio} processing, encompassing the broad categories of speech, music, and environmental sounds, is a vibrant research area that has a myriad of real-world applications. These applications range from voice-activated assistants like Siri and Alexa~\cite{hoy2018alexa,li2017acoustic} to transcription services~\cite{yoshioka2019advances}, and extend to telecommunication systems~\cite{luo2018tasnet} and hearing aids~\cite{latif2020speech}. Traditional audio processing systems were built on meticulously hand-crafted features and extensive linguistic knowledge~\cite{jelinek1998statistical}. 
Despite their effectiveness, these hand-crafted approaches often lacked scalability and struggled with the variability and complexity inherent in audio signals~\cite{gold2011speech}.
However, in the past decade, the field has experienced a significant paradigm shift with the emergence of data-driven methodologies~\cite{lecun2015deep,goodfellow2016deep,krizhevsky2012imagenet}. 
This progression towards data-centric techniques paves the way for systems that can learn to understand and interpret complex audio patterns directly from raw data~\cite{purwins2019deep,oord2016wavenet}.

$ $

However, these data-driven models, despite their prowess, typically perform well only for the specific tasks they are trained on and generally struggle with situations that deviate from their training environments.
Meanwhile, \textit{Large AI Models}, particularly Large Language Models (LLMs), have demonstrated outstanding accomplishments in almost every AI domain, reshaping how humans interact with machines~\cite{bommasani2021opportunities,zhao2023survey,teubner2023welcome}. These large models, characterised by their billions of parameters and training on massive datasets, have manifested emergent abilities to tackle a multitude of intricate tasks across various fields~\cite{wei2022emergent,schaeffer2023emergent,liu2023emergent}. Such capabilities have elevated AI algorithms to unprecedented levels of power and efficacy. In particular, the emergence of models such as ChatGPT and GPT-4 has rekindled discussions about the potential of artificial general intelligence~\cite{bubeck2023sparks, OpenAI-OpenAI-2023-GPT-4}. Unlike earlier learning-based models that were tailored for specific tasks, these large models boast versatility in addressing diverse tasks~\cite{brown2020language,alayrac2022flamingo}. Given their immense potential, these expansive AI models signify a new technological wave that promises a rich ecosystem of real-world applications and have already found extensive applications in various sectors such as vision~\cite{awais2023foundational,zhang2023vision}, language, health, education, robotics, and governance, among others.

\begin{table*}[!ht]
  \setlength\tabcolsep{5pt}
  \centering
  \caption{Comparison between this paper and other review articles concerning Foundation Models (FMs)/Large Language Models (LLMs) and/or Audio Signal Processing.}
  \resizebox{0.98\linewidth}{!}{
  \begin{tabular}{l|c|c|c|c|p{10.5cm}}
  \toprule[1pt]
    \textbf{Authors} & \textbf{Year} & \textbf{FM}& \textbf{Audio} & \textbf{Domain} & \textbf{Focus} \\
  \midrule
    Karita~\textit{et al.}~\cite{karita2019comparative} & 2019 & \large{\xmark} & \large{\cmark} & Speech &{Comprehensive} study to compare the performance of transformer and recurrent neural networks in numerous speech applications. \\
      
    Latif~\textit{et al.}~\cite{latif2023transformers} & 2022 & \large{\xmark} & \large{\cmark} & Speech &\textbf{First} survey paper of applications of transformer models in speech processing. \\

    Mehrish~\textit{et al.}~\cite{mehrish2023review} & 2023 & \large{\xmark} & \large{\cmark} & Speech &{Comprehensive} survey covering applications of deep learning in speech processing. \\   

    Latif~\textit{et al.}~\cite{latif2023survey} & 2023 & \large{\xmark} & \large{\cmark} & Speech &\textbf{First} survey paper of applications of reinforcement learning in audio processing. \\ \hline
    Bommasani~\textit{et al.}~\cite{bommasani2021opportunities} & 2022 & \large{\cmark} & \large{\xmark} & General &A comprehensive survey paper on the applications and risks of foundation models in diverse fields including language, vision, health, among others. \\
    Zhao~\textit{et al.}~\cite{zhao2023survey} & 2023 & \large{\cmark} & \large{\xmark} & General &\textbf{First} comprehensive survey paper on LLMs including their background, key findings in the literature, and mainstream techniques. \\
    Chang~\textit{et al.}~\cite{chang2023survey} & 2023 & \large{\cmark} & \large{\xmark} & General &Comprehensive review of these evaluation methods for LLMs, focusing on three key dimensions: what to evaluate, where to evaluate, and how to evaluate. \\
    Kaddour~\textit{et al.}~\cite{kaddour2023challenges} & 2023 & \large{\cmark} & \large{\xmark} & General & Identify several unsolved challenges of LLMs, provide an overview of their current applications, and discuss how the former constrain the latter. \\
    Wang~\textit{et al.}~\cite{wang2023aligning} & 2023 & \large{\cmark} & \large{\xmark} & General &\textbf{First} survey to provide an up-to-date review on the alignment process of LLMs. \\ \hline
    Gan~\textit{et al.}~\cite{gan2022vision} & 2022 & \large{\cmark} & \large{\xmark} & Vision & This survey categorises vision-language pre-training frameworks, covering various architectures, objectives, and downstream tasks. \\
    Zhang~\textit{et al.}~\cite{zhang2023comprehensive} & 2023 & \large{\cmark} & \large{\xmark} & Vision & Comprehensive review of the visually prompted foundation segmentation model, segment anything (SAM) and discusses potential downstream tasks. \\
    Zhang~\textit{et al.}~\cite{zhang2023vision} & 2023 & \large{\cmark} & \large{\xmark} & Vision & Survey of different vision-language pre-training network architectures, objectives, and downstream tasks and categorises vision-language pre-training frameworks. \\
    Awais~\textit{et al.}~\cite{awais2023foundational} & 2023 & \large{\cmark} & \large{\xmark} & Vision &Reviews vision and language foundational models focusing on their architecture types, training objectives, downstream task adaption, and their prompting designs with a broad coverage of their applications in a variety of visual tasks. \\ \hline
    Kasneci~\textit{et al.}~\cite{kasneci2023chatgpt} & 2022 & \large{\cmark} & \large{\xmark} & Education & Emphasise the potential of \textit{Large Models} models to enhance educational content, boost student engagement, and tailor individual learning experiences.\\
    Kung~\textit{et al.}~\cite{kung2023performance} & 2023 & \large{\cmark} & \large{\xmark} & Education & Assess ChatGPT's performance on the United States Medical Licensing Exam (USMLE), where it impressively achieved scores near the passing threshold without any dedicated specialised training. \\
    Qadir~\textit{et al.}~\cite{qadir2023engineering} & 2023 & \large{\cmark} & \large{\xmark} & Education & Review regarding promise and pitfalls of ChatGPT in engineering education. \\
    Rudoph~\textit{et al.}~\cite{rudolph2023chatgpt} & 2023 & \large{\cmark} & \large{\xmark} & Education & Examine the implications of technology for higher education, focusing on the future of learning, teaching, and assessment in the context of AI chatbots like ChatGPT. \\ \hline
    Moor~\textit{et al.}~\cite{moor2023foundation} & 2023 & \large{\cmark} & \large{\xmark} & Health & Identify potential applications for medical foundation models and outline specific technical capabilities and training data needed to enable them. \\
    Qiu~\textit{et al.}~\cite{qiu2023large} & 2023 & \large{\cmark} & \large{\xmark} & Health & Comprehensive review of \textit{Large AI Models} in health informatics including drug discovery, medical diagnosis and decision-making, medical imaging, medical informatics, medical education, public health, and medical robotics.  \\
    Wornow~\textit{et al.}~\cite{wornow2023shaky} & 2023 & \large{\cmark} & \large{\xmark} & Health & Reviews 84 foundation models using non-imaging EMR data, categorising their architectures, training sources, and applications. \\
    Zhang~\textit{et al.}~\cite{zhang2023challenges} & 2023 & \large{\cmark} & \large{\xmark} & Health & Survey of medical foundation models, from general vision to modality and task-specific ones, emphasising their challenges, opportunities, and uses. \\ \hline
    Hu~\textit{et al.}~\cite{hu2022protein} & 2022 & \large{\cmark} & \large{\xmark} & Comp.\ Bio. & Review the latest developments in \textit{Large Models} and \textit{Protein Large Models}, focusing on their architectures, pre-training methods, and prevalent protein databases. \\
    Tran~\textit{et al.}~\cite{tran2023survey} & 2023 & \large{\cmark} & \large{\xmark} & Comp.\ Bio. &Survey a number of representative embedding models for execution time, memory needs, and their ability to perform various tasks related to global properties for different protein sets \\ \hline

    Cyphert~\textit{et al.}~\cite{cyphert2021human} & 2022 & \large{\cmark} & \large{\xmark} & Law & Article delves into the ethical implications of integrating GPT-3 into legal practices. \\
    Sun~\textit{et al.}~\cite{sun2023short} & 2023 & \large{\cmark} & \large{\xmark} & Law &Survey of LLMs in legal tasks like judgement prediction and document analysis. Also highlights related legal challenges including privacy, bias, and transparency. \\ 
    Nay~\textit{et al.}~\cite{nay2023large} & 2023 & \large{\cmark} & \large{\xmark} & Law & Examines LLM's proficiency in tax law application, noting improvements in newer models compared to older ones. \\ \hline
    Yang~\textit{et al.}~\cite{yang2023foundation} & 2023 & \large{\cmark} & \large{\xmark} & Robotics &Explore applications of foundation models in practical decision-making using prompting, generative modeling, planning, and reinforcement learning. \\ \hline

    \rowcolor{orange!6} This paper & 2023 & \large{\cmark} & \large{\cmark} & Audio &\textbf{First} survey paper of applications of \textit{Large AI Models} in audio signal processing.\\ 
  \bottomrule[1pt]
  \end{tabular}
  }
  \label{tab:sum_TL}
\end{table*}

While large AI models have made remarkable advancements in the domains of language~\cite{paass2023foundation}, images~\cite{awais2023foundational}, and videos~\cite{zhao2023learning}, the audio arena has followed a more gradual trajectory. Nevertheless, recently, these large models have made significant strides in a variety of audio processing tasks, characterised by techniques that adeptly integrate audio data representations with traditional text token embeddings, equipping these large models with the capacity to interpret and manage a wide range of audio content~\cite{liu2023audioldm,fathullah2023prompting,li2023diverse}. 
Despite substantial progress and promising potential, the integration of large models into audio processing presents unique challenges and requires dedicated exploration.
This highlights the imperative for an all-encompassing survey centred on the application of these large models within the audio domain, encompassing speech, music, and other auditory facets. This paper aims to fulfil this requirement, providing an exhaustive overview of the methods, limitations, and future directions in this emerging field.
Specifically, our key contributions are as follows:

\begin{itemize}
  \item This is the first survey paper that comprehensively covers applications of \textit{Large AI Models} in the domain of audio signal processing, thereby 
  covering the recent progress in this emerging area.
  
  \item We also shed light on how \textit{Large AI Models} handle the distinct characteristics of audio processing and how they can be further enhanced to handle the complexities of spoken language. In particular, we cover the applications of these large models in the broad categories of speech and music.

  \item We discuss challenges, limitations, and potential directions for future research. Through this survey, we aim to provide a comprehensive understanding of the current landscape of large models in the realm of audio processing, thus paving the way for future innovations in this exciting area.
\end{itemize}

\begin{figure}[t]
\centering
\includegraphics[width=0.5\textwidth]{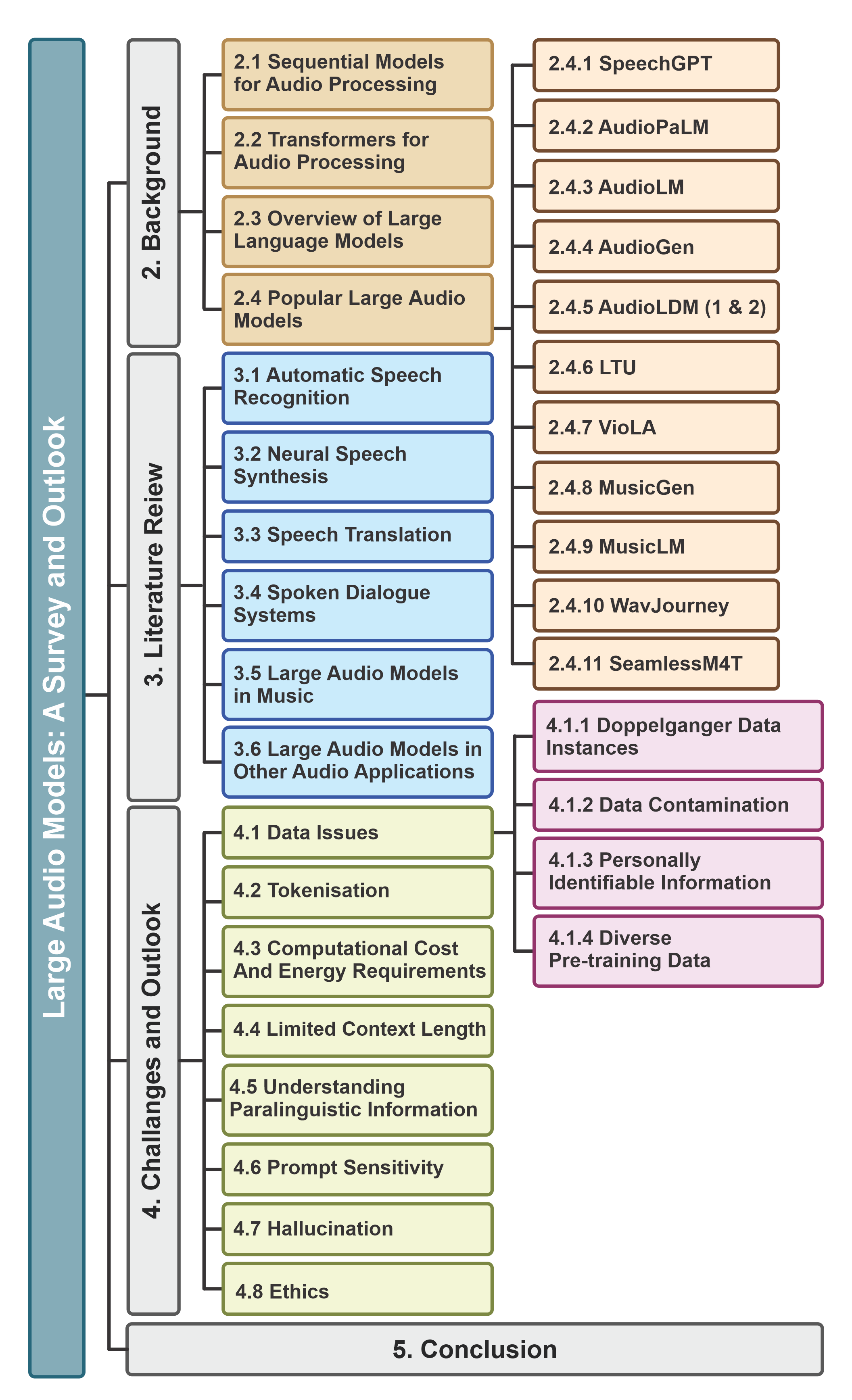}
\caption{Paper outline. }
\label{paperoutline}
\end{figure}

\textbf{\textit{Paper Organisation.}} The organisation of this paper is shown in Figure \ref{paperoutline}. Section~\ref{sec:background} provides insights into the applications of sequential models and transformers
within the audio processing sphere, while also briefly discussing large language models and the pivotal role of datasets in training expansive audio models. Section~\ref{sec:applications} provides a comprehensive overview of the applications of large AI models in the speech and music domains. Section~\ref{sec:challenges} discusses open
problems and charts potential avenues for future research. Finally, in Section ~\ref{sec:conclusion}, we summarise and conclude the paper.

\textbf{\textit{Related Surveys and Differences.} }
While several comprehensive surveys delve into the applications of deep learning for audio processing~\cite{purwins2019deep,deshpande2022ai,latif2023survey,liu2022audio}, including speech~\cite{karita2019comparative,latif2023transformers}, music~\cite{briot2017deep,ji2020comprehensive,moysis2023music}, and other categories~\cite{chachada2014environmental,miyazaki2019environmental}, none concentrate on the advent and deployment of LLMs in this field. 
Numerous surveys exist that cover the vast landscape of LLMs, each focusing on specific aspects or applications. Among these, the work by Zhao et al.~\cite{zhao2023survey} closely parallels ours, as it provides a broad overview of LLMs and related topics. Similarly, Mialon et al.~\cite{mialon2023augmented} turn their attention towards augmented language models, those with advanced reasoning capabilities and tool usage skills.
On a similar vein, Tornede et al.~\cite{tornede2023automl} explore LLMs in the context of Automated Machine Learning (AutoML) techniques, discussing existing methodologies and the challenges of using them to enhance LLM performance. Tang et al.~\cite{tang2023science} focus on techniques for detecting text generated by LLMs, while Chang et al.~\cite{chang2023survey} have examined the various ways to evaluate LLMs.
Additionally, there are a number of surveys dedicated to investigating the specialised applications of \textit{Large Models} in various fields such as vision~\cite{gan2022vision,zhang2023comprehensive,zhang2023vision,awais2023foundational}, education~\cite{kasneci2023chatgpt,kung2023performance,qadir2023engineering,dwivedi2023so,rudolph2023chatgpt}, healthcare~\cite{moor2023foundation,qiu2023large}, computational biology~\cite{tran2023survey,hu2022protein}, computer programming~\cite{xu2022systematic,wang2023codet5+}, law~\cite{sun2023short,nay2023large,trozze2023large,cyphert2021human}, or robotics~\cite{yang2023foundation,qin2023tool,schubert2023generalist} among others. On the other hand, \textbf{our survey stands apart in its exclusive focus on the applications of \textit{Large AI Models} in the realm of audio signal processing}, and fills an existing gap in the current body of research.
To round off our review, we provide a brief summary of the contributions of existing surveys in Table~\ref{tab:sum_TL}.

\section{Background} \label{sec:background}

In this section, we provide an overview of LLMs, beginning with a brief overview of sequential models and the difficulties they encounter while processing sequential data. Subsequently, we will probe the principal ideas that underpin the operation of large language models, emphasising the distinctive traits that equip these models to surpass traditional recurrent neural networks. Ultimately, we will examine the widely used large language models in the domain of audio processing.

\subsection{Sequential Models for Audio Processing} 
Initial applications of deep learning in the field of audio processing
primarily utilised versions of Convolutional Neural Networks (CNNs)~\cite{abdel2014convolutional}. However, the inability of these CNN-based methodologies to encapsulate the sequential essence of speech data was a substantial disadvantage. This shortcoming led to the inception of sequence-to-sequence (seq2seq) architectures, such as Recurrent Neural Networks (RNNs)~\cite{sutskever2014sequence} and Long Short-Term Memory Networks (LSTMs)~\cite{hochreiter1997long}, specifically engineered for handling sequential data. RNNs proved to be a suitable fit for sequential data given their ability to process extensive sequences incrementally, maintaining a constrained memory of preceding sequence components. A recent trend in research merges the unique strengths of both CNNs and RNNs. This involves using CNNs to derive audio features, which are then fed as input for RNN training. However, RNNs are known to suffer from the challenges of vanishing or exploding gradients.
\begin{figure*}[t]
\centering
\includegraphics[width=0.7\textwidth]{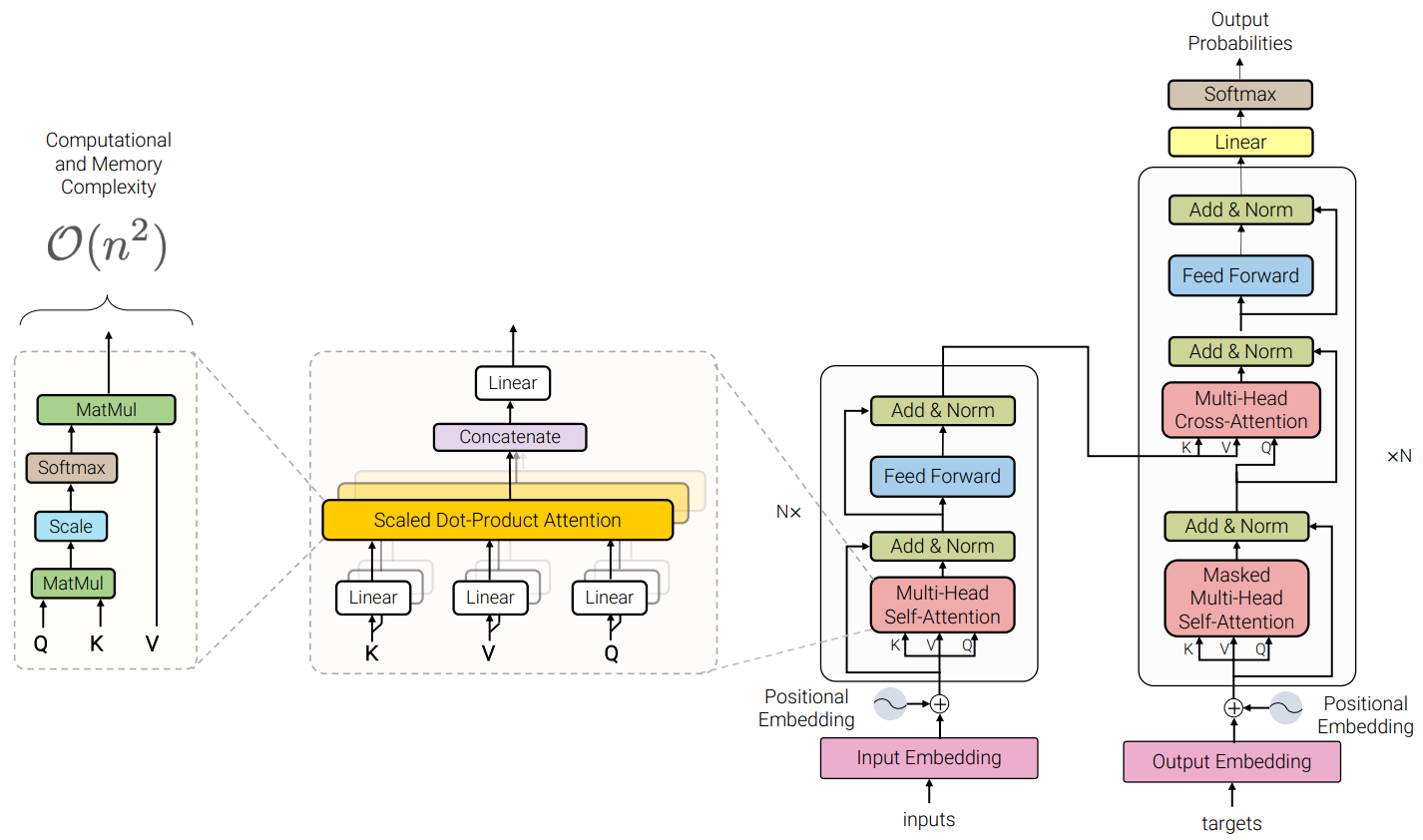}
\caption{Architecture of standard transformer, a fundamental building block of \textit{Large AI Models} (adapted from Vaswani et al.~\cite{vaswani2017attention} and Tay et al.~\cite{tay2022efficient}). It consists of encoder and decoder layers, both equipped with stacked self-attention and feed-forward components. The encoder derives hidden states from an input token sequence, and the decoder utilises these states alongside its own output token sequence to produce predictions.}
\label{trans_archi}
\end{figure*}
To combat this, LSTMs implement a gating mechanism alongside memory cells to regulate the information flow and mitigate issues related to gradients~\cite{yu2019review,salehinejad2017recent}. There have been various adaptations of LSTMs, such as Frequency-LSTM, Time-Frequency LSTMs, Bi-directional LSTMs, ConvLSTMs, and Stacked LSTMs, each proposed to cater to specific Speech Processing tasks. Despite their potency, seq2seq models have certain restrictions. For instance, they struggle to leverage parallel computing hardware efficiently and have difficulty in modelling long-term contexts due to their inherently sequential nature.

\subsection{Transformers for Audio Processing}

Transformers utilise self-attention mechanisms to capture temporal correlations from sequential data~\cite{vaswani2017attention}. This equips transformers with the ability to capture extensive temporal contexts while maintaining reduced computational complexity. Transformers employ self-attention layers to effectively capture distant relationships within input sequences, unlike traditional RNNs which struggle with such interactions. Self-attention also enables greater parallelisation compared to RNNs, allowing transformers to process speech sequences holistically without relying on past states. Vaswani et al.~\cite{vaswani2017attention} introduced two types of attention: scaled dot-product attention and multi-head attention. Additionally, positional encoding conveys information about token positions (see Figure~\ref{trans_archi}). These benefits have spurred significant interest in transformers across various AI domains~\cite{khan2022transformers,han2022survey,shamshad2023transformers,lu2022transformers,selva2023video,aleissaee2023transformers,wen2022transformers}, notably the audio community. This has given rise to diverse architectures such as Wav2Vec~\cite{schneider2019wav2vec}, Whisper~\cite{radford2023robust}, FastPitch~\cite{lancucki2021fastpitch}, MusicBERT~\cite{zeng2021musicbert}, and others~\cite{latif2023transformers,hernandez2022music,ji2023survey}.

Furthermore, transformers have not only revolutionised natural language processing and audio processing but have also paved the way for the development of LLMs that can understand, generate, and interact with human language and its underlying contexts in increasingly nuanced and sophisticated ways. Their remarkable ability to efficiently capture contextual dependencies and relationships within sequences has been instrumental in the creation of LLMs with billions of parameters, such as GPT-3. This breakthrough in capturing contextual information has extended beyond text generation to various modalities like speech and audio, giving rise to the emergence of \textit{Large Audio Models} that have transformed tasks such as speech recognition, emotion detection, and music generation. We discuss the \textit{Large Audio Model} in the next subsection. 


\begin{figure*}[t]
\centering
\includegraphics[width=1\textwidth]{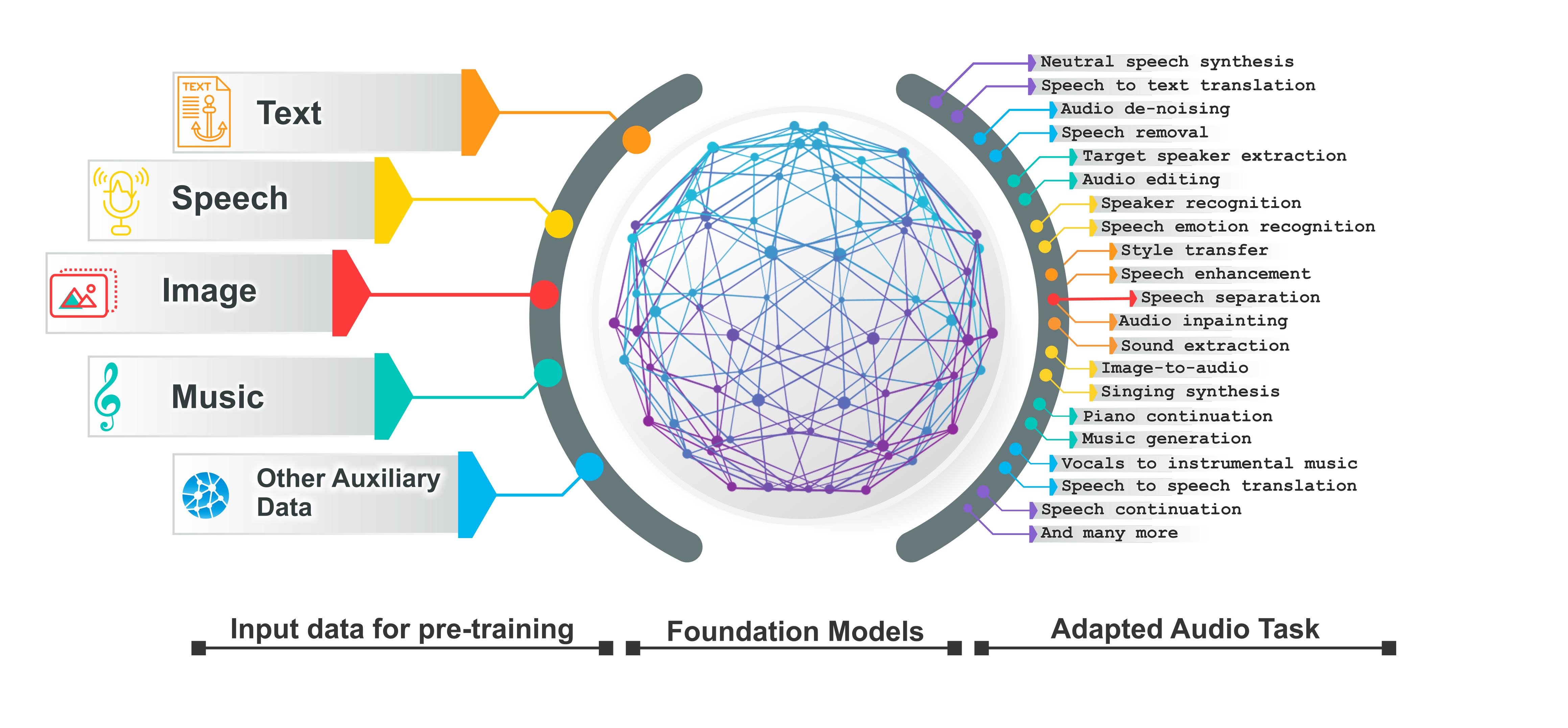}
\caption{Overview of \textit{Foundational Audio Models}: A \textit{Foundational Audio Model} aggregates information from diverse data modalities. Once trained, this model can be tailored to various downstream audio tasks.}
\label{largeaudiooverview}
\end{figure*}

\subsection{Overview of Large Language Models}

Investigations reveal that the act of scaling pre-trained language models (PLMs), either through enhancing the model size or expanding the data size, typically yields superior model performance on subsequent tasks, adhering to what is known as the scaling law~\cite{kaplan2020scaling}. Numerous investigations have probed the limits of performance by training increasingly larger PLMs, such as the GPT-3 model with 175 billion parameters and the PaLM model with 540 billion parameters. While the majority of scaling endeavours primarily focus on model size (preserving similar architectures and pre-training tasks), these expanded PLMs exhibit distinct characteristics compared to their smaller counterparts, such as BERT with 330 million parameters and GPT-2 with 1.5 billion parameters. They exhibit unexpected proficiency, referred to as emergent abilities, in tackling a variety of intricate tasks. For example, GPT-3 has demonstrated the ability to address few-shot tasks via in-context learning, a feat that GPT-2 struggles with. Hence, the term ``large language models (LLMs)" has been coined by the research community to describe these enlarged PLMs, and these models have garnered increasing interest. A notable example of an LLM application is ChatGPT, which adapts the GPT series LLMs for dialogue, showcasing exceptional conversational capabilities with humans. A significant surge in arXiv papers pertaining to LLMs can be observed following the launch of ChatGPT. 

\begin{figure*}[t]
\centering
\includegraphics[width=0.92\textwidth]{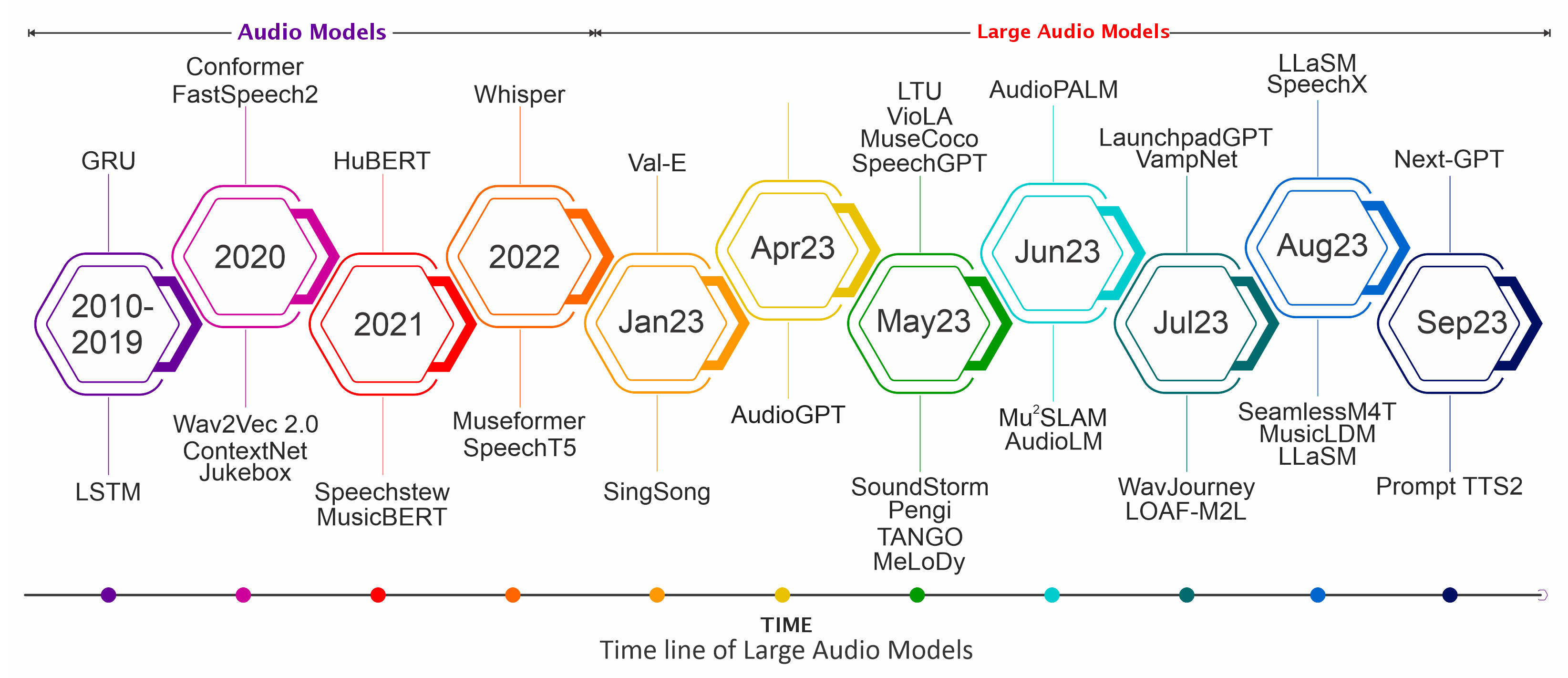}
\caption{Timeline of \textit{Large Audio Models}.}
\label{largeaudiotimeline}
\end{figure*}

Recently, GPT-4~\cite{openai2023gpt4} has been developed, which is a large-scale multimodal model that can accept image and text as input and produce text outputs. GPT-4 is capable of achieving human-level performance on some professional and academic benchmarks, including achieving a score around the top 10\% of test-takers in a simulated bar exam. Various other multimodal large language models are proposed by utilising multimodal information including visual, audio, and text. These LLMs are considered a crucial step towards Artificial General Intelligence (AGI). Most importantly, \textit{Large Audio Models} (see Figure \ref{largeaudiooverview}) attract significant interest from the research community to build LLMs that have intrinsic cross-modal conversational abilities and are capable of perceiving and generating audio or multimodal content. We also show a brief timeline for \textit{Large Audio Models} in Figure \ref{largeaudiotimeline}. In the next section, we cover popular \textit{Large Audio Models} and a summary of these models is presented in Table~\ref{llmsaudio}.

\subsection{Popular \textit{Large Audio Models}}
\label{largeaudiomodels}
In this section, we provide a brief overview of popular \textit{Large Audio Models}.

\subsubsection{SpeechGPT}

Zhang et al.~\cite{zhang2023speechgpt} proposed SpeechGPT, a large language model that has intrinsic cross-modal conversational abilities that allow it to generate multimodal content. The model is based on three significant elements: a discrete unit extractor, a large language modal, and a unit vocoder. They utilised Hidden-unit BERT (HuBERT)~\cite{hsu2021hubert} as a discrete unit extractor for the transformation of continuous speech to discrete units, the Meta AI LLaMA~\cite{touvron2023llama} model as LLM, and HiFi-GAN as a unit vocoder. The low availability of publicly available speech data compelled them to construct the SpeechInstruct, a speech-text cross-modal instruction-following dataset comprised of two parts cross-modal instructions and Chain-of-Modality Instruction. The training process of this model is broken down into three steps, Modality Adaptation Pre-training on unpaired speech data, Cross-modal Instruction Fine-Tuning, and Chain-of-Modality Instruction Fine-Tuning. They employ an unlabelled speech corpus to train the LLM in a next-token prediction task which empowers the Large Language Model (LLM) to effectively handle discrete units of modality. In the Cross-modal Instruction Fine-Tuning, they utilised the paired data to align speech and text. Subsequently, they applied the parameter-efficient Low-Rank Adaptation (LoRA) technique~\cite{hu2021lora} to perform fine-tuning. Consequently, they found the model to perform various tasks with correct output on different instructions. Although this model has shown remarkable cross-modal instruction recognition and speech dialogue abilities, it also has some limitations that can be listed as paralinguistic information, sequential response generation and context length limitation.
\subsubsection{AudioPaLM} 
\begin{figure*}[t]
\centering
\includegraphics[width=0.95\textwidth]{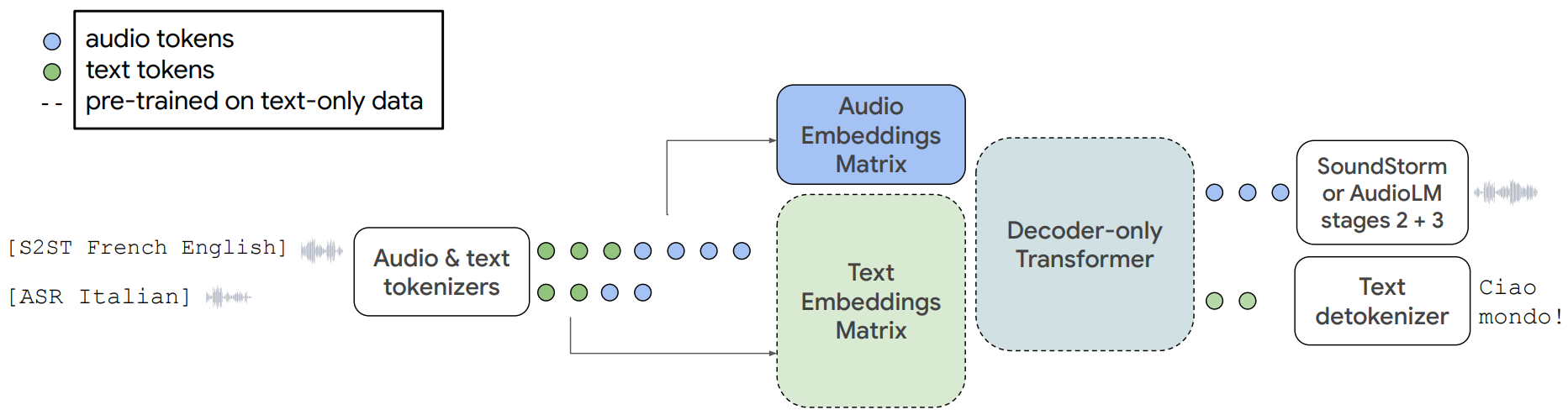}
\caption{Overview of the AudioPaLM model~\cite{rubenstein2023audiopalm} designed for speech-to-speech translation and automatic speech recognition. A pre-trained text-only model (denoted by dashed lines) is modified to incorporate an extended embedding matrix for new audio tokens. The overall structure remains consistent, accepting a combined sequence of text and audio tokens, and decoding either type. The subsequent stages of AudioLM or SoundStorm then revert audio tokens back to raw audio. Figure taken from~\cite{rubenstein2023audiopalm}.}
\label{AudioPaLMoverview}
\end{figure*}

Rubenstein et al.~\cite{rubenstein2023audiopalm} introduce a multimodal generative model called AudioPaLM (see figure \ref{AudioPaLMoverview}) for speech and text, capable of both understanding and generating speech. The model is built upon the foundation of PaLM~\cite{chowdhery2022palm} and PaLM-2~\cite{anil2023palm}, initially devised for text-only pre-training.
The model's training encompasses three primary stages: tokenisation of text and audio, modification of pre-trained text decoders, and transformation of the model's output into audio. They adopt token extraction techniques from raw audio~\cite{lakhotia2021generative,borsos2023audiolm}. Following token processing, the tokens are fed into a transformer decoder, which subsequently passes through an audio decoding process. They employ autoregressive techniques as in AudioLM~\cite{borsos2023audiolm}, as well as non-autoregressive approaches similar to~\cite{borsos2023soundstorm} to translate decoding tokens into audio. Their findings demonstrate improved ASR/AST performance with LLM size, and a single model is effectively trained across multiple tasks.

\subsubsection{AudioLM}
Borsos et al.~\cite{borsos2023audiolm} present the AudioLM framework, designed to facilitate high-quality audio synthesis while prioritising the preservation of long-term consistency, coherence, and uniformity across extended time spans. This framework is composed of three integral components: a tokeniser model, a decoder-only transformer, and a de-tokeniser model. Drawing from SoundStream~\cite{zeghidour2021soundstream}, w2vBERT~\cite{chung2021w2v}, the k-means quantiser for w2v-BERT embeddings, and decoder-only transformers, all of which have been trained on the extensive Libri-Light~\cite{kahn2020libri} English dataset encompassing 60,000 hours of speech data, the authors assembled these components. This amalgamation incorporates adversarial neural audio compression, self-supervised representation learning, and language modelling techniques. They have shown a comparison between the acoustic tokens from SoundStream and the semantic tokens extracted from a pre-trained w2v-BERT model on a speech dataset to show that these two types of tokens complement each other regarding enhancing phonetic discriminability and attaining high-quality rebuilding of the audio content. Through training on comprehensive raw audio waveform datasets, AudioLM acquires the proficiency to generate high-quality and logically coherent audio extensions from concise prompts. Converting input audio into a series of tokens, AudioLM approaches audio generation as a language modelling task.

\subsubsection{AudioGen}
Meta recently introduced AudioCraft, an extensive framework designed to facilitate a diverse range of generative audio tasks encompassing music generation, sound effects creation, and post-training compression using raw audio signals. This comprehensive framework consists of three essential components: MusicGen~\cite{kreuk2022audiogen}, AudioGen, and EnCodec. Both MusicGen and AudioGen incorporate independent autoregressive Language Models (LMs) tailored to operate with discrete audio representations in the form of tokens. In contrast, EnCodec is built upon neural networks.

AudioGen~\cite{kreuk2022audiogen}, a critical component of this framework, is an auto-regressive model that effectively addresses the challenge of generating audio while incorporating textual inputs. This model adopts a transformer-based architecture, functioning with discrete audio representations. 
The operational mechanism of this model can be distilled into two primary steps. Firstly, an auto-encoding method~\cite{zeghidour2021soundstream, li2021real} is employed to comprehend the discrete representation of raw, unprocessed audio. Subsequently, these acquired representations are employed to train the transformer language model. The transformer decoder language model is extended from the GPT2-like model, imbuing the entire system with an encoder-decoder configuration. Empirical evaluations underscore the model's commendable performance across both objective and subjective evaluation metrics, positioning it favourably in comparison to assessed baselines. Notably, the proposed methodology excels in generating audio continuations, adeptly navigating both conditional and unconditional scenarios.

\subsubsection{AudioLDM and AudioLDM 2}
AudioLDM~\cite{liu2023audioldm} is a text-to-audio generation framework with an encoder built on a contrastive language audio pre-trained (CLAP) model, and the latent diffusion model (LDM) for sound generation with audio embedding as input and text embedding as conditions. The CLAP model is pre-trained with datasets including LAION-Audio-630K, AudioSet, AudioCaps and Clotho. With the CLAP encoder, the training of the LDM does not require audio-text pairs any more, which is substantially different from the previous method such as AudioGen~\cite{kreuk2022audiogen} and DiffSound~\cite{yang2023diffsound}. As a result, a large number of audio clips (without the paired texts) could be used to train the LDM model, and this leads to a generation model capable of generating more diverse sounds with potentially better quality as compared with AudioGen and DiffSound. In addition, due to the operation in the latent space, the training of AudioLDM is much more efficient as compared with AudioGen and DiffSound, and only one GPU is required for training on the AudioCaps dataset. In addition, the AudioLDM model enables a number of other audio-related tasks to be performed in zero-shot fashion, such as text-guided super-resolution, inpainting, and style transfer. 
Built on the success of AudioLDM, the authors have created a more advanced model called AudioLDM 2~\cite{liu2023audioldm2}, which aims to develop a general audio representation method called "language of audio" (LOA), for speech, music and general sound effects. With this method, a single foundation model is learned with the same method and is able to generate high-quality speech, music and sound effects. The self-supervised learning method AudioMAE is used to convert any audio modality into the language of audio, With the LOA representation, the audio signal can be generated with a self-supervised learning process, with a LDM with LOA as conditions. This technique leverages the strengths of in-context learning, the pre-trained AudioMAE, and 
LDM. This method is shown to give state-of-the-art performance in text-to-sound generation.

\subsubsection{LTU}
Gong et al.~\cite{gong2023listen} present an audio model known as LTU (Listen, Think, and Understand), designed to perform audio classification and captioning tasks based on the OpenAQA-5M dataset, which comprises 5.6 million diverse audio samples. The training of LTU involves the creation of a novel dataset, OpenAQA-5M, by amalgamating eight datasets containing audio, questions, and answers. The architecture of the LTU model draws from various components, including an audio spectrogram transformer (AST)~\cite{gong2021ast} as the audio encoder, LLaMA~\cite{touvron2023llama} as the large language model (LLM) enhanced with Vicuna~\cite{chiang2023vicuna} instructions, Low-rank Adapter~\cite{hu2021lora}, and specific generation settings. To align the embedding dimensions with LLaMA, a pre-trained Audio Spectrogram Transformer is used alongside the CAV-MAE~\cite{gong2022contrastive} and fine-tuned on AudioSet-2M~\cite{gemmeke2017audio} for audio encoding.

During training, the authors maintained the LLaMA unchanged to minimise catastrophic forgetting~\cite{goodfellow2013empirical}. They focused solely on training the AST audio encoder, the audio projection layer, and the LoRA adapters. LLaMA underwent self-supervised pre-training on both natural language and programming language datasets, while Vicuna was fine-tuned using instructions generated by GPT models. The arbitrary initialisation of the audio projection layer led to training this component in conjunction with closed-ended classification and acoustic feature description tasks while keeping AST and LoRA adapters unaltered. Evaluation of LTU against a state-of-the-art model, CLAP, showcased its significant performance in audio-to-text tasks, achieving an average relative improvement of 23.1\%
across classification eight benchmarks. 

\subsubsection{VioLA}
Wang et al.~\cite{wang2023viola} introduce VioLA, a codec language model encompassing a multilingual multimodal auto-regressive transformer decoder-only network. This model exhibits proficiency in speech recognition, speech synthesis, and translation, covering speech-to-text (STT), text-to-speech (TTS), and machine translation (MT) tasks. VioLA is built upon VALL-E~\cite{wang2023neural} and VALL-E X~\cite{zhang2023speak}, which share TTS capabilities akin to GPT. The authors utilise an offline neural model, EnCodec, to convert speech waveforms into discrete tokens. This transformation enables speech representations to be treated as textual tokens, effectively leveraging a decoder-only model for adept optimisation of multimodal tasks. VIOLA is trained using multi-task learning strategies, encompassing ASR, MT, and TTS tasks. 
The results underscore VIOLA's effectiveness in addressing both single-modal and cross-modal tasks. Despite its versatility in numerous speech tasks, VIOLA is not without limitations. Its training relies solely on supervised data, neglecting the untapped potential of unsupervised data, including unlabelled speech and diverse text corpora. The model's scope encompasses in-context learning for speech synthesis tasks, but it does not encompass other speech processing (SP) tasks. Additionally, VIOLA currently lacks end-to-end capabilities in comprehensive speech-processing tasks.


\subsubsection{MusicGen}
MusicGen, a part of the AudioCraft framework~\cite{copet2023simple}, is a text-to-music generation language model (LM) that operates on discrete audio representations to generate music from provided text descriptions. This study introduces a model for generating coherent music based on text and melody conditions, with extensive objective and subjective evaluations. The architecture relies on an autoregressive transformer-based decoder~\cite{vaswani2017attention}, conditioned on textual and musical representations. Enodec~\cite{defossez2022high} is employed to encode audio into a continuous tensor. The model is trained on 20,000 instances of licensed music data and evaluated against MusicCaps benchmarks~\cite{agostinelli2023musiclm}, surpassing evaluated baselines in subjective assessments.

\begin{table*}[!ht]
\centering
\scriptsize
\caption{Some recent \textit{Large Audio Models}. ASR: automatic speech recognition, SS: speech synthesis, TTS: text to speech, ST: speech translation, SP: speech paralinguistics, SD: spoken dialogue system, code: official code release, {\large{\color{blue}{$\approx$}}}: will be released later.}
\label{llmsaudio}

\begin{tabular}{lllccccccc}
\toprule
\multirow{2}{*}{LLM/Paper} & \multirow{2}{*}{Train data} & \multicolumn{6}{c}{Tasks} \\ 
\cmidrule(r){3-8}
 & & ASR & TTS & ST & SP & SD & Others&Code \\
\midrule
SpeechGPT~\cite{zhang2023speechgpt}& \begin{tabular}[c]{@{}l@{}}Gigaspeech\\ Common Voice\\ LibriSpeech\\SpeechInstruct\end{tabular} & \large{\cmark} & \large{\cmark} & \large{\xmark} & \large{\xmark} & \large{\cmark} &-&\large{\xmark}\\\midrule

AudioPaLM~\cite{rubenstein2023audiopalm}& \begin{tabular}[c]{@{}l@{}}CoVoST2, CVSS\\ VoxPopuli ASR \\Common Voice\\ Conversational EsEn \\LibriSpeech\\YouTube ASR\\WMT/TED TTS\\PaLM MT TTS\end{tabular} & \large{\cmark} & \large{\cmark} & \large{\cmark} & \large{\xmark} & \large{\xmark} &Machine Translation&\large{\xmark}\\ \midrule


  

AudioLM~\cite{borsos2023audiolm}& Libri-Light & \large{\xmark} & \large{\xmark} & \large{\xmark} & \large{\xmark} & \large{\xmark} & \begin{tabular}[c]{@{}l@{}}Piano continuation\\Speech continuation\end{tabular}&\large{\xmark} \\ \midrule

LTU~\cite{gong2023listen}& OpenAQA-5M & \large{\xmark} & \large{\xmark} & \large{\xmark} & \large{\xmark} & \large{\xmark} & \begin{tabular}[c]{@{}l@{}}Audio classification\\ Audio captioning\\Summarisation\end{tabular}&\large{\color{blue}{$\approx$}} \\ \midrule

VIOLA~\cite{wang2023viola}& \begin{tabular}[c]{@{}l@{}}WenetSpeech\\Libri-Light\\LibriSpeech\\AI Challenger\\WMT2020\\EMIME\end{tabular} & \large{\cmark} & \large{\cmark} & \large{\cmark} & \large{\xmark} & \large{\xmark} & Machine translation &\large{\xmark}\\ \midrule

SpeechX~\cite{Xiaofei2023SpeechX}& \begin{tabular}[c]{@{}l@{}}LibriLight\\DNS challenge corpus\end{tabular} & \large{\xmark} & \large{\cmark} & \large{\xmark} & \large{\xmark} & \large{\xmark} & \begin{tabular}[c]{@{}l@{}}\hspace{1.5em}Noise suppression\\\hspace{1.5em}Speech removal\\\hspace{1.5em}Target speaker extraction\\\hspace{1.5em}Clean speech editing\\\hspace{1.5em}Noisy speech editing\end{tabular}&\large{\xmark} \\ \midrule


VALL-E~\cite{wang2023neural}& LibriLight & \large{\xmark} & \large{\cmark} & \large{\cmark} & \large{\xmark} & \large{\xmark} &- &\large{\xmark}\\ \midrule


Mu$^2$SLAM~\cite{cheng2023mu}& \begin{tabular}[c]{@{}l@{}}mC4 dataset\\ VoxPopuli, MLS, \\Babel, CoVoST \\ FLEURS.\end{tabular} & \large{\cmark} & \large{\xmark} & \large{\cmark} & \large{\xmark} & \large{\xmark} & Machine Translation &\large{\xmark}\\ \midrule

  
SoundStorm~\cite{borsos2023soundstorm}& LibriLight & \large{\xmark} &\large{\xmark} & \large{\xmark} & \large{\xmark} & \large{\cmark} &-&\large{\xmark} \\ \midrule

AudioGPT~\cite{huang2023audiogpt}& \begin{tabular}[c]{@{}l@{}}LibriTTS\\MUSTC\\CHiME4\\AudioSet\\AudioCaption\\and others\end{tabular} & \large{\cmark} & \large{\cmark} & \large{\cmark} & \large{\xmark} & \large{\cmark} & \begin{tabular}[c]{@{}l@{}}Style Transfer\\Speech Enhancement\\Speech Separation\\Mono-to-Binaural\\Audio Inpainting\\Sound Extraction\\Image-to-Audio\\Singing Synthesis\\and others\end{tabular}&\large{\cmark} \\ \midrule

  


Pengi~\cite{deshmukh2023pengi}& \begin{tabular}[c]{@{}l@{}}Clotho\\AudioCaps\\UrbanSound8K\\TUT 2017\\CREMA-D\\FSD50K\\and others\end{tabular} & \large{\cmark} & \large{\cmark} & \large{\cmark} & \large{\cmark} & \large{\xmark} & \begin{tabular}[c]{@{}l@{}}\hspace{2.5em}Audio Captioning\\\hspace{2.5em}Audio Question Answering\\\hspace{2.5em}Sound Sence Classification\\\hspace{2.5em}Music Analysis\\\hspace{2.5em}Instrument Classification\\\hspace{2.5em}Vocal Sound Classification\\\hspace{2.5em}and others\end{tabular}&\large{\color{blue}{$\approx$}} \\\hline


SeamlessM4T~\cite{seamlessm4t2023}& \begin{tabular}[c]{@{}l@{}}1 million hours \\of open speech \\audio data\end{tabular} & \large{\cmark} & \large{\cmark} & \large{\cmark} & \large{\xmark} & \large{\xmark} & \begin{tabular}[c]{@{}l@{}}\hspace{-0.5em}Machine Translation\\\hspace{-0.5em}{Speech,Text}-to-Text\\-Translation\end{tabular}&\large{\xmark} \\\hline

NExT-GPT~\cite{wu2023nextgpt}& \begin{tabular}[c]{@{}l@{}}T2M\\MosIT\end{tabular} & \large{\cmark} & \large{\cmark} & \large{\cmark} & \large{\xmark} & \large{\xmark} & \begin{tabular}[c]{@{}l@{}}\hspace{-0.5em}Text-to-Image\\Text-to-Video\\Text-to-Image\\\hspace{-0.5em}\end{tabular}&\large{\cmark} \\\hline

\bottomrule
\end{tabular}
\end{table*}

\subsubsection{MusicLM}
MusicLM~\cite{agostinelli2023musiclm} has the main idea of generating music from the textual description and it can generate high-quality music
at 24 kHz that has consistency over several minutes. It leverages the multi-stage autoregressive modelling of AudioLM~\cite{borsos2023audiolm} as the generative component and extends it to include text conditioning. It also uses MuLan~\cite{huang2022mulan}, a joint music-text model, to address the main challenge of paired data scarcity. 
The authors created a new hand-curated dataset, MusicCaps, which contains the 5.5k examples prepared by expert musicians. They trained the MusicLM to generate long and coherent music for textual descriptions of significant complexity. Based on the results, they showed that the MusicLM can generate up to 5-minute long clips and outperforms previous research in music quality as well as it adheres to the textual description. MusicLM inherits the limitations from MuLan, which makes the model misunderstand the negations which causes the model to not adhere to the temporal ordering described in the text.

\subsubsection{WavJourney}
WavJourney (see figure \ref{wavejourney}) is a method that uses LLMs to analyse text instructions and then connects a variety of audio models for compositional sound generation~\cite{liu2023wavjourney}. First, structured audio scripts are generated based on the text instruct using LLMs, and these scripts are organised in terms of their spatio-temporal relations. A script compiler is then used to convert the audio scripts into computer programs, which then calls for various acoustic models and operation functions in order to synthesise the audio content. This method offers a powerful creative tool for audio content generation, for a number of potential applications, including storytelling, science fiction, radio play, and education.

\subsubsection{SeamlessM4T}

\begin{figure}[t]
\centering
\includegraphics[width=0.48\textwidth]{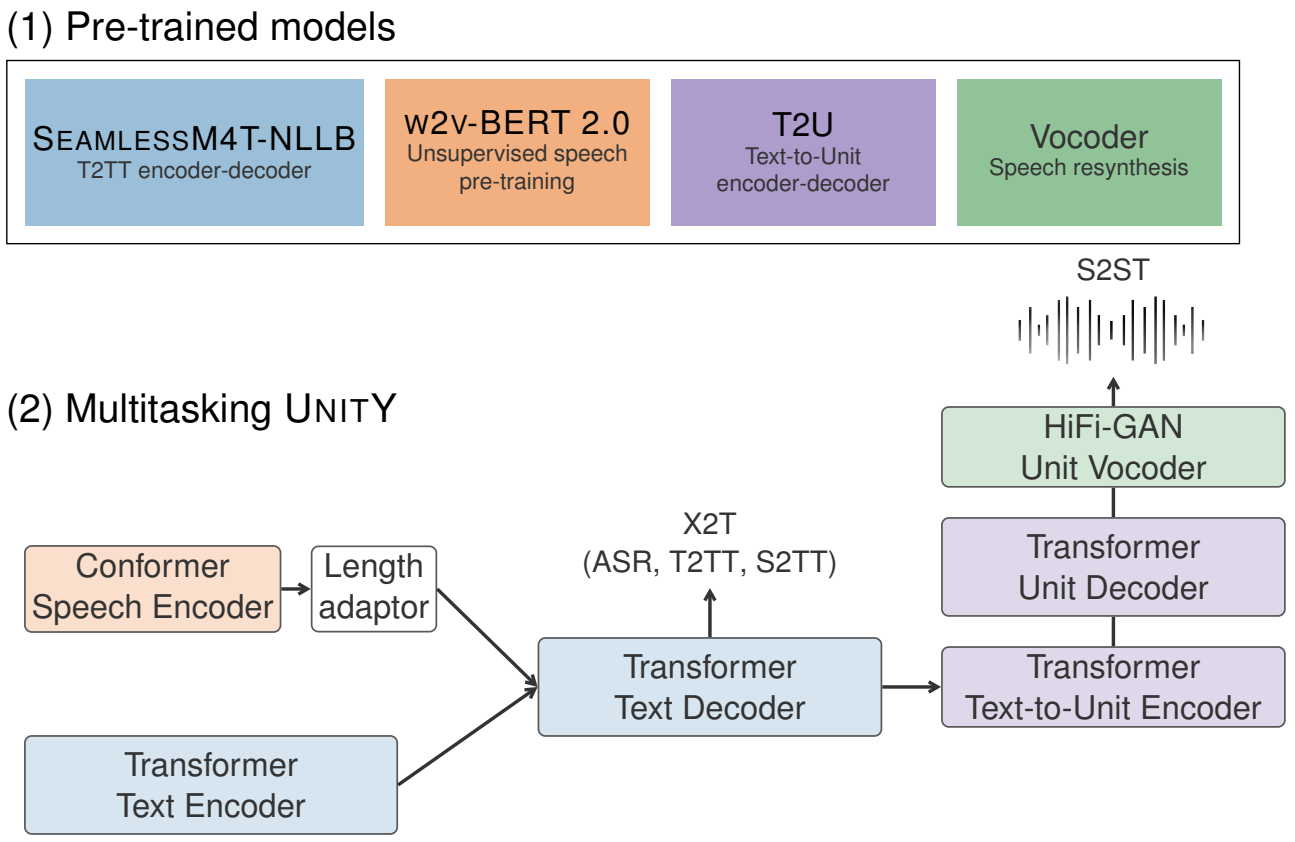}
\caption{Overview of the SeamlessM4T model: (1) Illustrates the pre-trained models employed during the fine-tuning of multitasking UnitY~\cite{inaguma2022unity}. (2) Depicts the multitasking UnitY structure, including its dual encoders, text decoder, T2U encoder-decoder, and accompanying vocoders for S2ST speech synthesis. Figure taken from~\cite{seamlessm4t2023}.}
\label{seamless}
\end{figure}
SeamlessM4T~\cite{seamlessm4t2023}, short for Massively Multilingual \& Multimodal Machine Translation (see figure 6 \ref{seamless}), offers a comprehensive solution for a wide range of translation tasks, spanning 100 languages. This model operates on the multitask UnitY architecture~\cite{inaguma2022unity}, facilitating the direct generation of translated text and speech, as well as supporting ASR and various translation modes. The architecture encompasses text and speech encoders, a text decoder, and a text-to-unit model, further strengthened by the self-supervised encoder, speech-to-text, text-to-text translation, and text-to-unit model pre-training. These components contribute to the conversion of decoded discrete units into speech through a multilingual HiFi-GAN unit vocoder~\cite{kong2020hifi}. Notably, the self-supervised speech encoder w2v-BERT 2.0 demonstrates improved training stability and representation quality, enabling the extraction of structural and semantic insights from multilingual speech. Alongside this, a text encoder trained across nearly 100 languages captures valuable text representations, enhancing the efficiency of multilingual translation tasks. 

\section{Literature Review} \label{sec:applications}
In this section, we extensively provide the literature review of \textit{Large Audio Models} in various tasks, including speech processing and music signal processing. For the evaluation of these tasks, various datasets are available and being used in audio processing research. In Table~\ref{audiodatasets}, we provide details of various public datasets used in the development of \textit{Large Audio Models}. For a comprehensive list of datasets, readers are referred to the GitHub page\footnote{https://github.com/EmulationAI/awesome-large-audio-models}. Below we cover various audio-related tasks using large audio models or LLMs. 
\begin{table}[!ht]
\scriptsize
\centering
\caption{List of audio datasets. ASR: automatic speech recognition, ST: speech translation, MT: machine translation, AC: audio classification, SED: sound event detection, AMG: affective music generation, MAG: music analysis and generation, MU: music understanding, SC: sound classification, SG: symphony generation, TTM: text to music, MT: music tagging. MAG: Music Arrangement Generation, MGR: Music Genre Recognition }. 
\begin{tabular}{lllll}
\hline
Title                             & Application          & Size                 & \begin{tabular}[c]{@{}l@{}}Multi-\\lingual\end{tabular}  & \begin{tabular}[c]{@{}l@{}}Public \\access\end{tabular}  \\ \hline
CommonVoice 11~\cite{ardila2019common}    & ASR              & 2508 hours              & \cmark     & \cmark      \\ \hline
Libri-Light~\cite{kahn2020libri}       & ASR              & 60000 hours             & \xmark      & \cmark      \\ \hline
Wenetspeech~\cite{zhang2022wenetspeech}    & ASR              & 10000 hours             & \xmark      &        \\ \hline
Gigaspeech~\cite{chen2021gigaspeech}     & ASR              & 50000 hours             & \xmark      & \cmark      \\ \hline
MuST-C~\cite{di2019must}                & \begin{tabular}[c]{@{}l@{}}ASR, MT \\and SLT\end{tabular}      & 3600 hours              & \cmark     & \cmark      \\ \hline
VoxPopuli~\cite{wang2021voxpopuli}      & ASR, S2ST           & 400k hours              & \cmark     & \cmark      \\ \hline
CoVoST~\cite{wang2020covost}                & ST              & 2880 hours              &   \cmark     & \cmark      \\ \hline
CVSS~\cite{jia2022cvss}                 & ST              & 3809 hours              & \cmark     & \cmark      \\ \hline
EMIME~\cite{wester2010emime}                 & ST              & -                  & \cmark     & \cmark      \\ \hline
Audiocaps~\cite{kim2019audiocaps}       & AC              & 46K audios           & -      & \cmark      \\ \hline
Clotho~\cite{drossos2020clotho}        & AC   & 

\begin{tabular}[c]{@{}l@{}}4981 audios \\24905 captions \end{tabular}

 & -      & \cmark      \\ \hline
Audio set~\cite{gemmeke2017audio}       & SED 
& \begin{tabular}[c]{@{}l@{}}5.8k hours \end{tabular} 

 & -      & \cmark      \\ \hline
EMOPIA~\cite{hung2021emopia}         & AMG   & \begin{tabular}[c]{@{}l@{}}387 piano \\solo sounds   \end{tabular}  & \cmark     & \cmark      \\ \hline
MetaMIDI~\cite{ens2021building}        & MCA              & 
\begin{tabular}[c]{@{}l@{}}436631 MIDI\\ files   \end{tabular}
        & -      & \cmark      \\ \hline
DALI2~\cite{meseguer2020creating}       & MU              & 7756 Songs              & -      & \cmark      \\ \hline
Million MIDI~\cite{zeng2021musicbert} & MU              & 100K Songs        & -      &        \\ \hline
Vggsound~\cite{chen2020vggsound}       & SC              & 
\begin{tabular}[c]{@{}l@{}}200k videos \end{tabular}

 & -      & \cmark      \\ \hline
FSD50K~\cite{fonseca2021fsd50k}        & SED & \begin{tabular}[c]{@{}l@{}}51197 sound\\ clips \end{tabular}           &       & \cmark      \\ \hline
Symphony~\cite{liu2022symphony}        & SG              & 
\begin{tabular}[c]{@{}l@{}}46359 MIDI \\files \end{tabular} 

          & -      & \cmark      \\ \hline
MusicCaps~\cite{agostinelli2023musiclm}    & TTM              & 
 \begin{tabular}[c]{@{}l@{}}5521 music-\\text pairs \end{tabular}

        & \xmark   & \cmark      \\ \hline
Jamendo~\cite{bogdanov2019mtg}                & MT               & 55525 tracks                   &       &  \cmark       \\ \hline

POP909~\cite{wang2020pop909}                &  MAG               &  \begin{tabular}[c]{@{}l@{}}909 songs,\\multiple piano\\arrangements\end{tabular}                   &       - &  \cmark       \\ \hline

FMA~\cite{defferrard2016fma}                &  MGR               &  \begin{tabular}[c]{@{}l@{}}106574 clips\end{tabular}                   &       - &  \cmark       \\ \hline

\end{tabular}
\label{audiodatasets}
\end{table}


\subsection{Automatic Speech Recognition (ASR)}
Automatic Speech Recognition (ASR) empowers machines to convert the spoken language into corresponding text sequences, comprising words or even sub-words. In ASR research, recurrent neural networks (RNNs) embedded with long short-term memory (LSTM)~\cite{hochreiter1997long} units are considered as core architecture until the transformers have been proposed~\cite{latif2023survey}. In contrast to RNNs, transformers can model temporal correlations within sequential data by utilising self-attention mechanisms~\cite{vaswani2017attention}. In addition, transformers offer the advantage of parallelising computations, enabling faster training of deeper models on larger datasets. Recently, language models have shown their power in capturing high-level, long-term patterns across different data types including text~\cite{brown2020language,chowdhery2022palm} and image~\cite{zhang2021vinvl,zhou2022conditional}, and speech~\cite{deng2022improving,higuchi2022bert,higuchi2023bectra}. This has also opened avenues for developing \textit{Large Audio Models} in the speech and audio domain. 

For instance, Wu et al.~\cite{wu2023decoder} introduced the concept of speech-LLaMA, a technique that involves seamlessly integrating acoustic embeddings into a text-based large language model to enhance translation capabilities. This integration empowers the language model to base its translation on acoustic cues. This model comprises three fundamental elements: a pre-trained text neural LM, an audio encoder, and a Connectionist Temporal Classification (CTC) 
compressor. They utilise LLaMA-7B~\cite{gao2023llama} as their text neural LM due to its flexibility. The CTC compressor, a pre-trained component, ensures the alignment of text and speech lengths. Simultaneously, an audio encoder facilitates the transformation of continuous speech vectors. 
Notably, the approach bypasses the conversion of speech into discrete tokens, instead directly mapping continuous speech representation into the LM's semantic space. This tailored architecture effectively accommodates acoustic embeddings within text-based language models, proficiently processing both acoustic embeddings and text cues to generate outputs that seamlessly integrate textual and acoustic insights. Kubo et al.~\cite{kubo2022knowledge} present a strategy to tackle this challenge through knowledge transfer from a neural network language model, initially pre-trained on text-only data. The core focus lies in transferring the inherent semantic understanding embedded within large-scale language model vectors. These vectors serve as implicit representations of linguistic aspects like part-of-speech and intent, holding potential as valuable cues for ASR decoders. The proposed approach's effectiveness manifests in the form of reduced error rates, achieved without introducing extra computational complexities during the decoding phase.

Ling et al.~\cite{ling2023adapting} explore a methodology involving the use of pre-trained LLM for fully formatted End-to-End (E2E) ASR transcriptions. Their model architecture demonstrates flexibility in integrating a speech decoder with a pre-trained LLM, offering both encoder-decoder and decoder-only configurations. Drawing from a rich dataset of 75,000 hours of diverse formatted audio data spanning multiple domains, their approach remains highly adaptable. The composability of their model allows for the seamless integration of a speech encoder into a pre-trained LLM, featuring either an encoder-decoder or decoder-only structure. In the encoder-decoder-based LLM approach, a pre-trained LLM is harnessed, utilising its text tokeniser for speech recognition.
 Their training strategy encompasses three loss functions: CTC, Cross-Entropy (CE), and Masked Language Modeling (MLM), facilitating the acquisition of transcription knowledge from both textual and speech-text data. In the case of the decoder-only LLM approach for speech recognition, Ling et al.\ leverage the LoRA adapter to integrate it with the pre-trained LLM. This adaptation effectively minimises trainable parameters by updating pairs of decomposition matrices while preserving the original weights unaltered. For the encoder-decoder-based LLM, the Z-Code++ model~\cite{he2022z} serves as the text encoder and decoder. Conversely, the decoder-only LLM approach employs the GPT-2 model~\cite{radford2019language} as the decoder-based LLM. For performance comparison, the authors conduct thorough evaluations on a range of datasets, analysing the outcomes of five distinct models in their study.

In written text, meaningful sentence boundaries are often indicated by punctuation marks. However, this clear demarcation is lacking in spoken real-world utterances. To tackle this issue, Huang et al.~\cite{huang2023semantic} devised a strategy to extract punctuation insights from a bidirectional teacher language model (LM) trained on written and punctuated text. Their approach involves a comparison between their segmenter, distilled from the LM teacher, and another segmenter derived from an acoustic-pause-based teacher utilised in previous research. The evaluation of both segmenters took place within a streaming ASR pipeline. The incorporation of their segmenter led to a 3.2\% relative reduction in word error rate (WER) and a significant 60 ms reduction in median end-of-segment latency during a YouTube captioning task. In the previous section, we discussed AudioPaLM~\cite{rubenstein2023audiopalm}, a substantial \textit{Large Audio Models} designed to encompass both speech comprehension and generation. With a unified vocabulary bridging text and speech through a limited set of discrete tokens and a basic markup description of tasks, this model facilitates training a single decoder-only model for various tasks, including ASR. 
Evaluation efforts delved into ASR performance across multiple datasets, including CVSS, VoxPopuli ASR, CommonVoice 11, Conversational EsEn, and Youtube ASR datasets. The results highlight the model's competitive performance across these diverse datasets. In a different study, Huang~\cite{huang2022sentence} introduces strategies for curating language modelling data to enhance the recognition of rare words without compromising overall performance. These strategies demonstrate substantial impact, leading to an enhanced language model achieving a noteworthy up to 24\% relative reduction in WER for sentences containing rare words. Importantly, this enhancement in rare word recognition is achieved without causing any adverse impact on the overall WER.

Fathullah et al.~\cite{fathullah2023prompting} delve into extending the practicality of LLMs by directly incorporating a compact audio encoder, thus enabling them to perform speech recognition tasks. This approach for constructing multilingual speech recognition systems relies on Decoder-only LLMs conditioned on audio sequences. 
The underlying concept revolves around utilising large language models to capture sequences of embeddings, irrespective of their modality. By utilising a conformer-based audio encoder to generate embedding sequences and validating them through simple CTC loss training, this study leverages the LLaMA-7B~\cite{touvron2023llama} model with LoRA~\cite{hu2021lora} adaptation. The Multilingual LibriSpeech (MLS) dataset derived from LibriVox~\cite{pratap2020mls}, encompassing 50,000 hours of speech recordings in 08 different languages, serves as the basis for evaluation. The study's observations emphasise the alignment between audio embeddings and text, as well as the significance of audio encoder strides and size. Zhuo et al.~\cite{zhuo2023lyricwhiz} introduce LyricWhiz, a multilingual Automatic Lyrics Transcription (ALT) method designed for zero-shot scenarios across diverse lyrics transcription datasets, including unique genres like rock and metal. GPT-4, a large language model, serves as the annotator, while the Whisper speech recognition model~\cite{radford2023robust} assists in audio transcription. Leveraging the MTG-Jamendo dataset with 55,000 audio songs in various languages, the model requires no training and undergoes direct testing on multiple datasets, including Jamendo~\cite{stoller2019end}, Hansen~\cite{hansen2012recognition}, MUSDB~\cite{schulze2021phoneme}, and DSing~\cite{dabike2019automatic}. This combined approach not only transcribes lyrics in multiple languages but also contributes to reducing the WER in English. Furthermore, the model generates an extensive multilingual publicly available lyrics dataset based on MTG-Jamendo, offering a human-annotated subset for noise level estimation and evaluation.

\begin{table}[!ht]
\centering
\scriptsize
\caption{Average normalised WER comparison on Fleurs dataset for ASR. Where n is the number of languages.}
\begin{tabular}{llll}
\hline
\multirow{2}{*}{Model} & \multirow{2}{*}{Size} & \multicolumn{2}{c}{WER$\downarrow$}                                                         \\ \cline{3-4} 
            &            & \multicolumn{1}{l|}{\begin{tabular}[c]{@{}l@{}}Fleurs\\ (n=77)\end{tabular}} & \begin{tabular}[c]{@{}l@{}}Fleurs-54\\ (n=54)\end{tabular} \\ \hline
Whisper-Large-v2    & 1.5B         & \multicolumn{1}{l|}{41.7}                          & 43.7                            \\ \hline
MMS-L61        & 1.0B         & \multicolumn{1}{c|}{-}                            & 31.0                            \\ \hline
MMS-L1107       & 1.0B         & \multicolumn{1}{c|}{-}                            & \textbf{18.7}                       \\ \hline
SeamlessM4T-Medium   & 1.2B         & \multicolumn{1}{l|}{\textbf{21.9}}                      & 22.0                            \\ \hline
SeamlessM4T-Large   & 2.3B         & \multicolumn{1}{l|}{23.1}                          & 23.7                            \\ \hline
\end{tabular}
\label{benchmark}
\end{table}

In summary, recent advancements in leveraging LLMs or designing large audio models for speech-related tasks demonstrate the growing potential of combining linguistic and acoustic insights. Table \ref{llmsaudio} provides a concise overview of the various studies and their contributions. These studies highlight diverse strategies, from incorporating audio encoders to enhancing rare-word recognition and multilingual transcription. Table \ref{benchmark} compares the performance of SeamlessM4T with state-of-the-art ASR models including Whisper and MMS, which shows that \textit{Large Audio Model} considerably improves the ASR performance. As the field continues to evolve, these innovations underscore the capacity of language models to bridge the gap between speech and text, opening up new avenues for more efficient and effective solutions in speech processing and understanding.

\subsection{Neural Speech Synthesis}
Neural speech synthesis also referred to as Neural text-to-speech (TTS), is considered an important area of research with the aim of generating human-like speech from the text. Traditional TTS systems have complex architecture by encompasses intricate components including acoustic frontends, duration models, acoustic prediction models, and vocoder models. This complexity of TTS systems has recently been overcome with the advent of deep end-to-end TTS architectures. These systems possess the capacity to generate convincingly realistic speech by being trained on pairs of text and audio. Popular TTs models include Tacotron ~\cite{wang2017tacotron}, Deep Voice model~\cite{arik2017deep}, and Clarinet~\cite{ping2018clarinet}, and many other \cite{latif2023transformers}. These models produce Mel-spectrograms from textual inputs, which are subsequently employed for speech synthesis by vocoders like Griffin-Lim~\cite{griffin1984signal}, WaveNet~\cite{vanwavenet}, and Waveglow~\cite{prenger2019waveglow}. Lately, transformers become popular structures in TTS by showing improved performance and accelerated training \cite{latif2023transformers}. 

More recently, \textit{Large Audio Models} have become popular in solving problems in TTS research. Various studies either utilise LLMs or develop \textit{Large Audio Models} to show their effectiveness in the TTS domain. For example, Kakouros et al.~\cite{kakouros2023investigating} explore the concept of word surprisal as a potential factor enhancing prosody in speech synthesis. Word surprisal, a linguistic and NLP concept, quantifies the information conveyed by a word within a sentence or language model context. Their primary focus was investigating the interplay between word surprisal derived from LLMs and their capacity to capture prosodic prominence in both human and synthesised speech. Their study employed GPT-2 models and GPT-J, an open-source and open-access alternative to GPT-3, utilising the LJ Speech corpus as their dataset to assess surprisal rates in the textual content. The authors identified tokens and common sequences within the text, serving not only to satisfy the model's dictionary requirements but also to reduce the model's dictionary size and manage out-of-vocabulary (OOV) words. 
Hassid et al.~\cite{hassid2023textually} introduced TWIST, an innovative approach to training SpeechLMs that employs a warm-start strategy with a pre-trained textual LLM. This method capitalises on the shared characteristics between text and semantic tokens by initialising a decoder-only audio generator with the pre-trained weights of a text-based language model. Through a comprehensive combination of automated and human evaluations, TWIST consistently showcases superior performance compared to a cold-start SpeechLM across various aspects. Based on the results, the authors emphasise the importance of both model and dataset scale in enhancing the effectiveness of SpeechLMs.
 
Wang et al.~\cite{wang2023neural} trained a neural codec language model (called VALL-E) using discrete codes obtained from a readily available neural audio codec model. They approached TTS as a conditional language modelling task, differing from prior methods that treated it as a continuous signal regression. In the pre-training phase, they significantly expanded the TTS training dataset to 60,000 hours of English speech, a several-hundred-fold increase over existing systems. 
Experimental results show that VALL-E outperforms the leading zero-shot TTS system, particularly in terms of speech naturalness and speaker similarity. Additionally, results indicate that VALL-E effectively maintains emotional nuances and acoustic characteristics from the provided acoustic prompt during synthesis. VALL-E X, introduced in~\cite{zhang2023speak}, is designed for cross-lingual speech synthesis. It builds upon the foundation of VALL-E~\cite{wang2023neural} and is trained to predict acoustic token sequences in the target language speech using both source language speech and target language text as cues. VALL-E X inherits robust in-context learning capabilities, enabling its application in zero-shot cross-lingual text-to-speech synthesis and speech-to-speech translation tasks. Experimental results showcase its ability to generate high-quality speech in the target language using just a single speech utterance in the source language as input. 
This preservation of the unseen speaker's voice, emotion, and acoustic context is a prominent aspect of VALL-E X's performance.

Kharitonov et al.~\cite{kharitonov2023speak} presented a multi-speaker TTS SPEAR-TTS with two features of minimum data requirement for training and speech synthesis maintaining voice characteristics of a previously unseen speaker using a 3-second-long voice example. In particular, they integrate BART/T5-style pertaining~\cite{lewis2020bart,raffel2020exploring} with back translation~\cite{sennrich2016improving} to substantially decrease the quantity of parallel supervision necessary for training SPEAR-TTS. To control the voice employed by SPEAR-TTS during utterance generation, they utilise an illustrative prompting mechanism similar to textual language models~\cite{brown2020language}. They utilise LibriLight data as a source of training data and show that SPEAR-TTS attains a character error rate (CER) that is comparable with state-of-the-art techniques by only using 15 minutes of parallel data. Moreover, it matches the naturalness and acoustic quality of ground-truth speech, as assessed through subjective tests. VioLA~\cite{wang2023viola} (discussed in Section \ref{largeaudiomodels}) is a multilingual multimodal auto-regressive transformer decoder-only network that presents promising results in TTS. 
Their findings showcase a notable enhancement of 2.0\% in speaker similarity, a reduction of 14.6\% in WER, and an improvement in speech naturalness by 0.02.

Maiti et al.~\cite{maiti2023speechlmscore} introduced an autonomous evaluation approach known as SpeechLMScore, aimed at assessing generated speech samples using speech-language models. This unsupervised speech evaluation metric leverages a pre-trained language model to gauge the similarity between synthesised speech and natural human speech. The authors harnessed pre-trained models from GSLM~\cite{lakhotia2021generative} through fairseq \footnote{https://github.com/facebookresearch/fairseq} and employed the VoiceMOS challenge dataset~\cite{huang2022voicemos}, which encompasses speech from diverse sources. Encoding was accomplished using the pre-trained tokeniser HUBERT-BASE-LS960H~\cite{hsu2021hubert}, complemented by a k-means clustering model for quantisation. This combination of Hubert features and corresponding clustering models facilitated the development of uLM within GSLM with heightened efficiency. The model was exclusively trained with a dataset, eliminating the need for extensive human-evaluated data. In the context of an extensive dataset and larger model, the system was configured into four layers: SpeechLMScore (Pre), SpeechLMScore (LSTM), SpeechLMScore (LSTM)+rep, and SpeechLMScore (Large).

Wang et al.~\cite{wang2023lm} presented an LM-based approach named LM-VC for zero-shot voice transformation. This model draws inspiration from AudioLM and HuBERT. LM-VC is structured in two stages: 1) coarse acoustic modelling and 2) fine acoustic modelling. Within the LM-VC architecture, three distinct LMs are employed: a masked prefix LM (MPLM), an external LM (ELM), and a prefix LM (PLM). Leveraging the benefits of HuBERT and SoundStream, the model capitalises on separate sequences of semantic tokens and acoustic tokens. 
For training, the authors utilised LibriTTS and an internal dataset for both their model and SoundStream. Testing was conducted on a selection of 500 pairs from EMIME, VCTK, and CMU Arctic datasets. The model demonstrated efficiency in terms of the proximity of generated speech to natural speech and its similarity with the original speaker. Wang~\cite{wang2023assessing} proposed a method to assess phrase breaks utilising pre-trained language models and LLMs. The approach encompasses two key components: evaluating phrase breaks within speech and conducting a comprehensive analysis of each pause or break position. BERT was chosen for pre-training due to its vast training data and contextual understanding of word relationships. Additionally, the authors investigated the potential of ChatGPT for zero-shot and few-shot phrase break assessments. The authors used LJ speech data for pre-training and curated a dataset comprising 800 samples from diverse Chinese ESL learners, categorised as poor, fair, great, and humanly validated.  They demonstrate that the dependency of pre-trained language models has significantly decreased, leading to improved performance based on the results.



\begin{table}[!ht]
\centering
\caption{Neural speech synthesis comparison using LibriSpeech dataset.}
\begin{tabular}{|llll|}
\hline
\multicolumn{1}{|l|}{\multirow{2}{*}{Model}} & \multicolumn{1}{c|}{\multirow{2}{*}{WER $\downarrow$}} & \multicolumn{1}{c|}{\multirow{2}{*}{SPK$\uparrow$}} & \multirow{2}{*}{SMOS$\uparrow$} \\
\multicolumn{1}{|l|}{}            & \multicolumn{1}{c|}{}           & \multicolumn{1}{c|}{}           &            \\ \hline
\multicolumn{4}{|c|}{Speech-to-Speech Systems}                                                        \\ \hline
\multicolumn{1}{|l|}{GSLM~\cite{lakhotia2021generative}}          & \multicolumn{1}{l|}{12.4}         & \multicolumn{1}{l|}{0.126}        & \multicolumn{1}{c|}{-} \\ \hline
\multicolumn{1}{|l|}{AudioLM}        & \multicolumn{1}{l|}{6.0}         & \multicolumn{1}{c|}{-}          & \multicolumn{1}{c|}{-} \\ \hline
\multicolumn{4}{|c|}{TTS Systems}                                                               \\ \hline
\multicolumn{1}{|l|}{YourTTS~\cite{casanova2022yourtts}}        & \multicolumn{1}{l|}{7.7}         & \multicolumn{1}{l|}{0.337}    & 3.45±0.09       \\ \hline
\multicolumn{1}{|l|}{VALL-E}         & \multicolumn{1}{l|}{5.9}         & \multicolumn{1}{l|}{0.580}        & 4.38±0.10       \\ \hline
\multicolumn{1}{|l|}{VALL-E-continual}    & \multicolumn{1}{l|}{3.8}         & \multicolumn{1}{l|}{0.508}        & \multicolumn{1}{c|}{-} \\ \hline
\multicolumn{1}{|l|}{GroundTruth}      & \multicolumn{1}{l|}{2.2}         & \multicolumn{1}{l|}{0.754}        & 4.5±0.10        \\ \hline
\end{tabular}
\label{TTScomp}
\end{table}

We cover various recent papers on \textit{large audio models} or LLMs for neural speech synthesis. Table \ref{TTScomp} presents the benchmark results on the LibriSpeech dataset. Here WER is calculated on the generated speech and speaker similarity score (SPK) is calculated using the speech pairs from the same speaker in the test set. Human evaluation is performed to calculate SMOS on 40 speakers on LibriSpeech test-clean with a 3-second enrolled recording. Results show that VALL-E considerably outperforms other state-of-the-art models. In summary, speech synthesis has greatly benefited from complementing \textit{Large Audio Models} with acoustic-phonetic linguistic models as shown by the systems deployed in Table \ref{tab:TTS} summarise recently proposed \textit{Large Audio Models} evaluated on speech synthesis tasks.

\begin{table}[!ht]
\tiny
\caption{Summary of recent \textit{Large Audio Models} evaluated on text to speech (TTS) task.}
\centering
\begin{tabular}{|l|c|cccc|}
\hline
\multicolumn{1}{|c|}{\multirow{2}{*}{Model/Paper}} & \multirow{2}{*}{Dataset}                                           & \multicolumn{4}{c|}{Evaluations}                                                                                     \\ \cline{3-6} 
\multicolumn{1}{|c|}{}                             &                                                                    & \multicolumn{1}{c|}{MOS-P}       & \multicolumn{1}{c|}{MOS-Q}       & \multicolumn{1}{c|}{MOS-S}       & MOS         \\ \hline
\multirow{2}{*}{Mega-TTS:}                         & VCTK                                                               & \multicolumn{1}{c|}{4.32 ± 0.11} & \multicolumn{1}{c|}{4.27 ± 0.09} & \multicolumn{1}{c|}{4.27 ± 0.10} & -           \\ \cline{2-6} 
                                                   & LibriSpeech                                                        & \multicolumn{1}{c|}{4.21±0.17}   & \multicolumn{1}{c|}{4.08 ± 0.17} & \multicolumn{1}{c|}{3.90 ± 0.18} & -           \\ \hline
Mega-TTS 2                                         & LibriSpeech                                                        & \multicolumn{1}{c|}{4.11 ± 0.12} & \multicolumn{1}{c|}{4.15 ± 0.10} & \multicolumn{1}{c|}{4.02 ± 0.15} & -           \\ \hline
PromptTTS 2                                        & \begin{tabular}[c]{@{}c@{}}Multilingual\\ LibriSpeech\end{tabular} & \multicolumn{1}{c|}{-}           & \multicolumn{1}{c|}{-}           & \multicolumn{1}{c|}{-}           & 3.88 ± 0.08 \\ \hline

FoundationTTS                                        & \begin{tabular}[c]{@{}c@{}}Combined LibriTTS,\\ VCTK, and internal\end{tabular} & \multicolumn{1}{c|}{-}           & \multicolumn{1}{c|}{-}           & \multicolumn{1}{c|}{-}           & 3.98 ± 0.08 \\ \hline

\end{tabular}
\label{tab:TTS}
\end{table}


\subsection{Speech Translation (ST)}
Speech Translation (ST) involves the conversion of spoken speech from the source language into the target language. ST systems are typically categorised into two main groups: cascaded systems and end-to-end systems. Cascaded ST systems comprise an automatic speech recognition (ASR) component and a machine translation (MT) component. 
In contrast, end-to-end ST systems aim to optimise a single model that directly translates the spoken utterance into the target language. 
Various studies have explored methods and techniques to improve both cascaded ST systems~\cite{matusov2005integration} and end-to-end ST systems~\cite{berard2016listen}. In end-to-end ST systems, transformer-based models~\cite{latif2023transformers} have played a significant role in addressing various challenges. 
Recently, the use of \textit{Large Audio Models} is becoming increasingly popular in speech translation and showing promising results.

In the landscape of recent advancements, the introduction of SeamlessM4T~\cite{seamlessm4t2023} (as outlined in Section \ref{largeaudiomodels} ) stands out as a groundbreaking multimodal translation model, denoted as Massively Multilingual \& Multimodal Machine Translation (SeamlessM4T). The scope of this model is all-encompassing, spanning a multitude of translation tasks such as speech-to-speech, speech-to-text, text-to-speech, text-to-text, and ASR. Its capabilities extend across a wide linguistic panorama, spanning up to 100 languages. SeamlessM4T utilises the SeamlessAlign corpus, a monumental multimodal translation dataset totalling 470k hours, facilitated by the SONAR sentence embedding space adept at capturing both speech and text nuances. Notably, SeamlessM4T sets a new translation benchmark, exhibiting a 20\% BLEU improvement over prior direct speech-to-text methods on the Fleurs dataset. 

Dong et al.~\cite{dong2023polyvoice} introduced the innovative Poly Voice framework, which hinges upon a versatile language model (LM) proficient in speech-to-translation (S2ST) capabilities. This framework comprises two pivotal components: a translation language model and a speech synthesis language model. The former operates as a decoder-only model, while the latter involves discrete units. The translation model further delves into speech-to-unit translation (S2UT), effectively converting audio into language-specific units, while the speech synthesis model, identified as unit-to-speech (U2S), undertakes the task of generating translated speech while preserving the original speaker's style. The authors use HuBERT for semantic unit extraction (S2UT), while the U2S component employs the VALL-E X approach to execute speech synthesis. Additionally, SoundStream is enlisted to acquire embeddings of audio tokens. The training process involves multiple datasets spanning various domains encompassing ASR (LibriLight(En), In-house (Zh)), MT (In-house), and S2S (GigaSpeech, Wenet Speech). In the evaluation phase, two established benchmarks, namely EMIME and CVSS, are utilised to gauge speech and translation quality, providing comprehensive insights into the framework's performance.

As outlined in models, Rubenstein et al.~\cite{rubenstein2023audiopalm} proposed a multimodal generative model called AudioPaLM for speech based on the foundation of PaLM~\cite{chowdhery2022palm} and PaLM-2~\cite{anil2023palm}. The model can perform multiple tasks including Speech to Speech Translation (S2ST). 
 To build PaLM MT TTS, they employed PALM-2 for translating YouTube, CommonVoice, and Babel~\cite{gales2017low}. Consequently, after the training (described earlier) their model outperformed the baselines in AST and S2ST. Building upon the previous discussion, 
Wang et al.~\cite{wang2023viola} proposed VioLA, a language model encompassing a decoder-only transformer network which is multilingual and multimodal based on an auto-regressive approach that exhibits proficiency in speech-related tasks with the capability of speech translation. The model is based on VALL-E~\cite{wang2023neural} and VALL-E X~\cite{zhang2023speak}, an offline neural model, and EnCodec. The training procedure of the model has been previously outlined in the model section \ref{largeaudiomodels}. As a result, they found the model achieving improvement in BLUE scores.

The integration of speech and language training is confronted by challenges stemming from data and GPU requirements, as well as the inherent distinctions between spoken and textual information. Le et al.~\cite{le2023comsl} introduce ComSL, a novel speech-language model formulated through a composite architecture that harnesses the power of pre-trained speech and language models. This strategy optimises data utilisation for tasks involving spoken language. Specifically, ComSL incorporates cross-modality learning into transfer learning and concurrently applies these mechanisms within a multi-task learning framework for downstream tasks. Notably, ComSL demonstrates efficacy in end-to-end speech-to-text translation assignments. It achieves a remarkable new state-of-the-art average BLEU score of 31.5 on the multilingual speech-to-English text translation task across 21 languages, as assessed on the publicly available CoVoST2 dataset. Wu et al.~\cite{wu2023speechgen} conducted pioneering research that explores the application of prompt tuning to enhance speech-language models for a wide array of generation tasks. This innovative approach is implemented within a unified framework known as SpeechGen, characterised by its capacity to harness around 10 million trainable parameters. This cohesive framework holds significant promise, delivering increased efficiency and efficacy. The authors evaluated SpeechGen across three speech-related tasks, including speech translation, and demonstrated promising results.

In summary, the landscape of speech translation is evolving rapidly, with a growing focus on bridging the gap through innovative \textit{Large Audio Models}. The studies discussed in this section, as outlined in \ref{largeaudiomodels}, underscore the progress in this field. From leveraging large language models like AudioPaLM to tackle multilingual speech translation to the development of VioLA, a versatile language model proficient in speech-related tasks, these advancements hold the potential to revolutionise the accuracy and naturalness of translated speech. As the demand for seamless communication across languages continues to rise, these models offer a promising path forward in achieving enhanced speech translation capabilities.

\subsection{Spoken Dialogue Systems}

Spoken dialogue systems (SDSs) have garnered significant attention in the audio processing community due to their versatile applications in customer service and goal-oriented human-computer interactions. These systems encompass key components such as speech recognition, intent recognition, a knowledge base and/or database backend, a dialogue manager, language generation, and speech synthesis~\cite{zue2000conversational}. Within the architecture of SDSs, the dialogue manager plays a pivotal role in making action selections based on observed events~\cite{latif2023survey}. Researchers have effectively demonstrated how RNNs and transformers can be employed to optimise action selection, adeptly modelling the dynamic nature of spoken dialogue using fully or partially observable Markov Decision Processes. However, transformers have recently emerged as a superior alternative to RNNs to optimise the action selection process within SDSs~\cite{vlasov2019dialogue,wang2020vd}. By leveraging their self-attention mechanism, transformers have demonstrated exceptional capabilities in modelling dynamic dialogue system scenarios~\cite{zhang2023dynalogue}.

This evolution has led to numerous studies that harness the power of transformers to enhance (spoken) dialogue systems. While text-based dialogue systems can be trained directly on extensive text data~\cite{gao2018neural,serban2018survey}, a large number of SDSs have relied on user simulations for training due to the scarcity of real training dialogues available for both training and evaluation purposes ~\cite{ FanL20}. The integration of transformers into SDSs presents a promising avenue for improving dialogue management, offering the potential to better comprehend user inputs, context, and preferences, thus leading to more effective and natural interactions. Furthermore, the advances made in LLMs, such as those used in chat systems, and \textit{Large Audio Models}, have also paved the way for transformative changes in spoken dialogue systems. By leveraging knowledge acquired from pre-trained LLMs and \textit{Large Audio Models}, current and future SDSs may no longer require training from scratch or in isolation from other models. Instead, SDSs can inherit knowledge from large language/audio/multimodal models to bootstrap their input features, finetune or guide their behaviour, and potentially improve their performance. While direct usage of LLMs for task-oriented dialogue systems has shown to underperform in comparison with task-specific models, careful application is required for LLMs to be useful---as shown by~\cite{HuEtAl2023llms} in automated scoring of user simulations, \cite{HudecekDusek2023} in dialogue state tracking, and~\cite{WeiEtAl2023llms} in data collection via prompt engineering. This could be especially beneficial to task-oriented spoken dialogue systems with small or modest datasets. These models bring a new dimension of understanding and contextuality to conversations, not only in text but also in audio and visual interactions, opening doors to even more sophisticated and dynamic interactions between humans and machines.

However recent developments on LLMs-based dialogue systems are mostly text-based, and their application to spoken dialogue systems, audio-based conversational AI and their applications largely remain unexplored. A few exceptions using reasonably \textit{Large Audio Models} include dialogue generation from raw audio excluding text processing in their pipeline~\cite{ GSDLM2022}, dialogue policy learning from textual and audio features for task-oriented dialogues ~\cite{asier_ieee_taslp2023}, and open and closed-domain
dialogue generation \cite{cherakara2023furchat}. Other works on audio-based dialogue generation from audio features using \textit{Large Audio Models} include SpeechGPT~\cite{zhang2023speechgpt}, SoundStorm ~\cite{borsos2023soundstorm}, AudioGPT~\cite{huang2023audiogpt}, and dGSLM\cite{tacl_a_00545}. Further, recent studies such as ANGIE~\cite{LiuWZDWL022}, Multimodal-GPT\cite{gong2023multimodalgpt}, and Large Multimodal Models~\cite{li2023large} have integrated either vision and LLMs or video and audio~\cite{ LiLZFZ21} for training multimodal dialogue systems. Those efforts will be potentially transferable to LLM-based robot dialogue systems. 

The studies above have provided valuable insights regarding the potential applications and capabilities of large language and audio models within the context of SDSs. In the next years, we should expect a lot more influence of LLMs applied to SDSs---including speech and audio data (among others) in their learnt representations instead of only text---in order to improve their performance and acceptance by end users in a wide range of tasks. But additional aspects will have to be taken into consideration such as scarcity of audio and multimodal dialogue data (with representative amounts of transcriptions and annotations), safety of dialogues, and evaluations in real scenarios beyond simplified datasets.

\subsection{Large Audio Models in Music}
Deep Learning (DL) models find widespread application in content generation, spanning various domains such as images, text, and music. Particularly in music generation, DL's adaptability shines, allowing it to learn from a wide array of musical sources and enabling the creation of diverse genres. This sets it apart from conventional methods~\cite{briot2020deep}. The advent of transformers, renowned for their capacity to grasp intricate patterns and interdependencies in sequential data, has brought about a revolution in music generation. By comprehending long-range dependencies, harmonies, and subtleties, transformers have transformed the landscape of music generation~\cite{yu2022museformer,kumar2023words}. This transformation owes much to the self-attention mechanism within transformers, which incorporates a global context during the music composition process, resulting in outputs that are more coherent and sophisticated~\cite{ji2023emomusictv}. Moreover, the emergence of \textit{Large Music Models}, with transformers as a fundamental block, has further elevated music generation. These models harness the power of large AI models to craft music that resonates with human emotion and creativity, thus shaping the landscape of music composition in innovative and compelling ways. Below, we provide an extensive overview of \textit{Large Audio Models} with a focus on music signal processing.

Several prominent \textit{Large Audio Models} have emerged to advance the realm of music generation. For instance, Garcia et al.\cite{garcia2023vampnet} proposed a novel method known as VAMPNET. This approach hinges on masked acoustic token modelling and incorporates parallel iterative decoding. The foundational principles of VAMPNET are inspired by the Masked Global Information Tokeniser (MaskGIT) methodology\cite{chang2022maskgit}. The authors constructed their audio tokenizer using the Descript Audio Codec (DAC)~\cite{kumar2023high} and leveraged a multilayer bidirectional transformer~\cite{shaw2018self} for token prediction. The model was trained on an extensive dataset comprising 7,97,000 music tracks. For the assessment of audio quality, the researchers employed two key metrics: multiscale Mel-reconstruction and Fréchet Audio Distance (FAD)~\cite{kilgour2018fr}. The results of their experiment reveal that the model holds promise in generating music, particularly when short-loop recordings are used as input. 

Similarly, Ghosal et al.~\cite{ghosal2023text} introduced TANGO, an innovative approach designed for generating music from text inputs by leveraging the capabilities of FLAN-T5~\cite{chung2022scaling}. The TANGO architecture consists of three primary components: a text-prompt encoder, a latent diffusion model (LDM)~\cite{liu2023audioldm}, and a mel-spectrogram Variational Auto-Encoder (VAE)\cite{kingma2013auto}. The FLAN-T5-LARGE (780M) model, a pre-trained large language model, serves as the audio encoder, converting text inputs into encoded representations. The LDM is integral to the architecture. In addition, audio-VAE compresses mel spectrogram representations, while the audio synthesis stage employs HiFi-GAN~\cite{kong2020hifi} to transform mel spectrograms produced by the VAE decoder into audio. Experiments for this text-to-audio generation leveraged the AudioCaps dataset~\cite{kim2019audiocaps}, which consists of 45,438 audio samples. 
To assess the quality of the audio generated from the mel-spectrograms produced by the VAE decoder, Ghosal et al. utilised the vocoder introduced by Liu et al.~\cite{liu2023audioldm}. Benchmarking TANGO against established models such as DiffSound~\cite{yang2023diffsound}, AudioGen~~\cite{kreuk2022audiogen}, and AudioLDM~\cite{liu2023audioldm} highlighted its superiority in the domain of music generation from text input.

\begin{figure*}[t]
\centering
\includegraphics[width=0.7\textwidth]{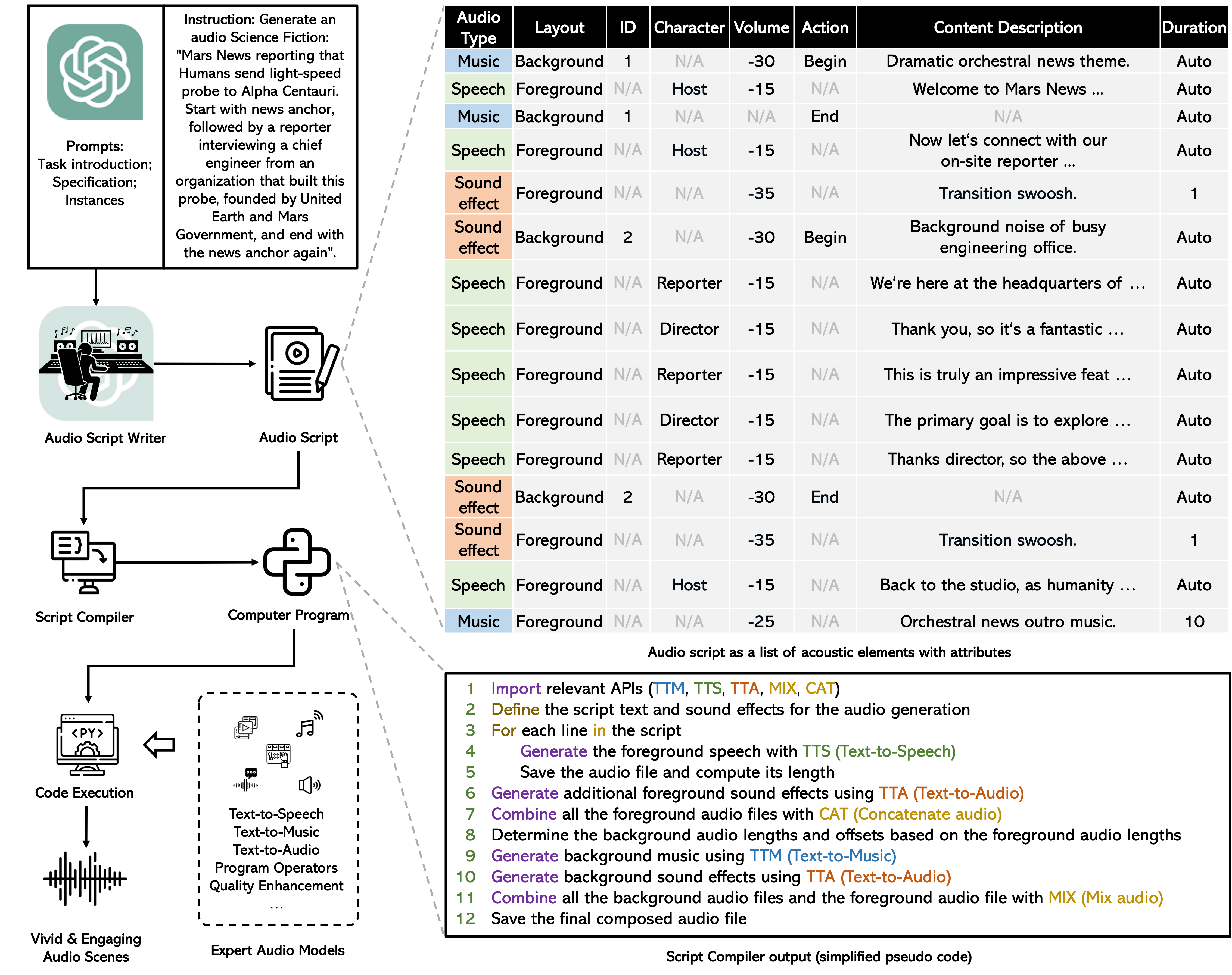}
\caption{An overview of WavJourney~\cite{liu2023wavjourney}: Initially, the LLM serves as an audio scriptwriter, offering users an interactive and clear representation of audio. This audio script is then processed by a script compiler and run akin to a computer program, utilising specialised audio generation models for execution. Figure taken from~\cite{liu2023wavjourney}.}
\label{wavejourney}
\end{figure*}

As presented in Section \ref{largeaudiomodels}, Liu et al.~\cite{liu2023wavjourney} presented WavJourney, a pioneering approach for generating comprehensive audio content—spanning speech, music, and sound effects—from textual story narrations.
WavJourney utilises the potential of LLMs to streamline the creation of audio compositions. Through the use of text-based directives, it possesses the capability to craft audio narratives that effortlessly blend elements such as speech, music, and sound effects, all without requiring any additional training. WavJourney showcases its versatility across a range of real-world applications, spanning from Science Fiction to educational content, radio plays, and AudioCaps. Chen et al.~\cite{chen2023musicldm} introduced the MusicLDM model, tailored for music generation from textual inputs. This model's conceptual foundation is anchored in Stable Diffusion\cite{rombach2022high}, the contrastive language audio pre-training model (CLAP)~\cite{wu2023large}, and the Hifi-GAN vocoder~\cite{kong2020hifi}. In particular, the approach of MusicLDM has significant similarities with AudioLDM~\cite{liu2023audioldm}. In its approach, audio data undergoes processing via an audio encoder, whereas text data is funnelled through a text encoder; both pathways are designed to extract semantic embeddings. Following this, the audio is converted into a mel-spectrogram and subsequently introduced into a variational autoencoder~\cite{kingma2013auto} to obtain the latent representation of audio. During the training phase, the CLAP model~\cite{wu2023large} is refined on the Audiostock dataset, which features 9,000 music tracks for training and an additional 1,000 tracks reserved for testing. This comprehensive training and evaluation dataset fortifies MusicLDM's proficiency, empowering it to craft music that aligns with the provided text input.

Wu et al.~\cite{wu2022exploring} devised a transformer-based model adept at producing music from textual descriptions. Concurrently, they introduced the Textune dataset, comprising 282,270 text-tune pairs that traverse a myriad of musical genres. 
They utilise various pre-trained checkpoints, purposed for natural language processing, to lay the groundwork for their approach. These checkpoints spanned from a randomly initialised encoder to established models like BERT~\cite{devlin2018bert}, GPT-2~\cite{radford2018improving}, and both BART base and large models~\cite{lewis2019bart}. The integration of a diverse array of pre-trained checkpoints, combined with transformer-based architectures, underpins their model's proficiency in translating text to music. The Textune dataset's richness further refines their transformer-centric methodology for deriving music from the text. Drawing a parallel, Huang et al.~\cite{huang2023noise2music} demonstrated that utilising LLMs to craft descriptive musical sentences can enhance the synthesis of text-conditioned music when utlised with a diffusion model.

Donahue et al.~\cite{donahue2023singsong} introduced SingSong, a novel method for generating instrumental music tailored to complement specific vocal inputs. At its core, this method focuses on generative modelling to produce instrumental music that harmoniously aligns with the provided vocals. 
To train the model, the authors utilised a dataset consisting of 1 million audio samples, which translates to roughly 46,000 hours of music. SingSong's foundation is anchored in AudioLM~\cite{borsos2023audiolm}. Throughout the training phase, the model is fed source-separated vocals as input, while the instrumental tracks act as the target. To counteract any artefacts in the vocals that might originate from instrumental segments, white noise was incorporated into the input. Semantic codes were mined using a pre-trained w2v-BERT model~\cite{chung2021w2v}, and coarse acoustic codes were extracted using a pre-trained SoundStream codec~\cite{zeghidour2021soundstream}. In addition, they used T5~\cite{raffel2020exploring}, an encoder-decoder transformer~\cite{vaswani2017attention}, to predict the output codes. The decoding operation was facilitated by SoundStream. Results show that SingSong stands out as a potent tool for producing instrumental music that aligns with given vocals, thereby enriching the music generation landscape with added creativity and depth.

Ou et al.~\cite{ou2023loaf} addressed the challenge of singability in generating lyrics. Specifically, their approach bridges the singability gap using a new method to produce singable lyrics by concurrently training on wording and formatting within the Melody-to-Lyric process (LOAF-M2L).
The foundation model for their approach is a transformer encoder-decoder model, built on the BART base architecture~\cite{lewis2019bart}. For training, they employed the DALI v2 dataset~\cite{meseguer2020creating} that aligns the lyric text with the melody and supplemented it with a text-only corpus from Kaggle~\cite{kaggledataset}. The model processes input from two primary channels: length prompts and specific note information from the melody. Each note detail undergoes embedding, and the resultant embeddings are aggregated into a note embedding vector. This vector is then merged with the length embedding, feeding into the encoder. The encoder's resultant embedding is pivotal, facilitating syllable stress classification into primary stress, secondary stress, and unstressed categories. Additionally, word importance labels are categorised into nonstopwords, secondary-important nonstopwords, and stopwords. In essence, the tools and strategies formulated by Ou et al. converge to create a melody-to-lyric generation model, uniquely adept at crafting singable lyrics, enhancing the music experience.

Lam et al.~\cite{lam2023efficient} introduced MeLoDy (M for music; L for Large Models; D for diffusion), an innovative audio music generation approach centred around a Language guided diffusion model~\cite{lam2023efficient}. The core concept involves leveraging a dual-path diffusion (DPD) model for acoustic modelling and a language model for semantic modelling. For the purpose of learning representation, the authors used MuLan~\cite{huang2022mulan}, Wav2Vec2-Conformer~\footnote{https://huggingface.co/docs/transformers/model\_doc/wav2vec2-conformer}, and VAE. These components were integrated into the MeLoDy framework to facilitate the effective generation of audio music. The dual-path diffusion model is employed to efficiently model coarse and fine acoustic information simultaneously. 
Their model was trained on an extensive dataset that contained 6.4 million audio samples, equivalent to 257,000 hours of audio content. The generation of music captions was facilitated by ChatGPT. To support the semantic modelling aspect, they trained a 429.5M parameter LLaMA~\cite{touvron2023llama} model with 24 layers, 8 heads, and 2048 hidden dimensions, and a scale comparable to MusicLM~\cite{agostinelli2023musiclm}. Ultimately, MeLoDy offers a fusion of LM-guided diffusion and dual-path diffusion models, effectively enhancing the generation of audio music by harmonising semantic and acoustic elements.

Lu et al.~\cite{lu2023musecoco} introduced MuseCoco (Music Composition Copilot), a robust framework designed to generate music based on textual descriptions. Their model operates in two distinct phases: the comprehension of text-to-attributes and the subsequent generation of music based on those attributes. In the initial text-to-attributes extraction phase, the authors employed BERT\textsubscript{large}~\cite{devlin2018bert} to tokenise and transform the input text into meaningful music attributes. This process allows the textual descriptions to be translated into a format suitable for music generation. To train attribute-to-music conversion, they used a variety of MIDI datasets, including  Million MIDI Dataset ~\cite{zeng2021musicbert}, EMPOIA~\cite{hung2021emopia}, MetaMidi~\cite{ens2021building}, POP909~\cite{wang2020pop909}, Symphony~\cite{liu2022symphony}, and an internal dataset named Emotion-gen. 
For the music generation phase, the REMI-like technique~\cite{huang2020pop} was employed to convert MIDI into token sequences. The Linear Transformer~\cite{pmlr-v119-katharopoulos20a} then served as the foundation for synthesising music from these attributes. In a comprehensive comparison, Lu et al. tested MuseCoco against GPT-4 and BART-base~\cite{wu2022exploring} using ABC notation as a shared format. To ensure a fair evaluation, the ABC notation music was converted to MIDI using the music21 tool\footnote{http://web.mit.edu/music21}. Overall, MuseCoco's two-fold approach of text-to-attribute understanding and attribute-to-music generation demonstrates its ability to produce musically coherent compositions from textual input, introducing a fresh avenue for creativity and collaboration in music composition.
\begin{figure}[t]
\centering
\includegraphics[width=0.49\textwidth]{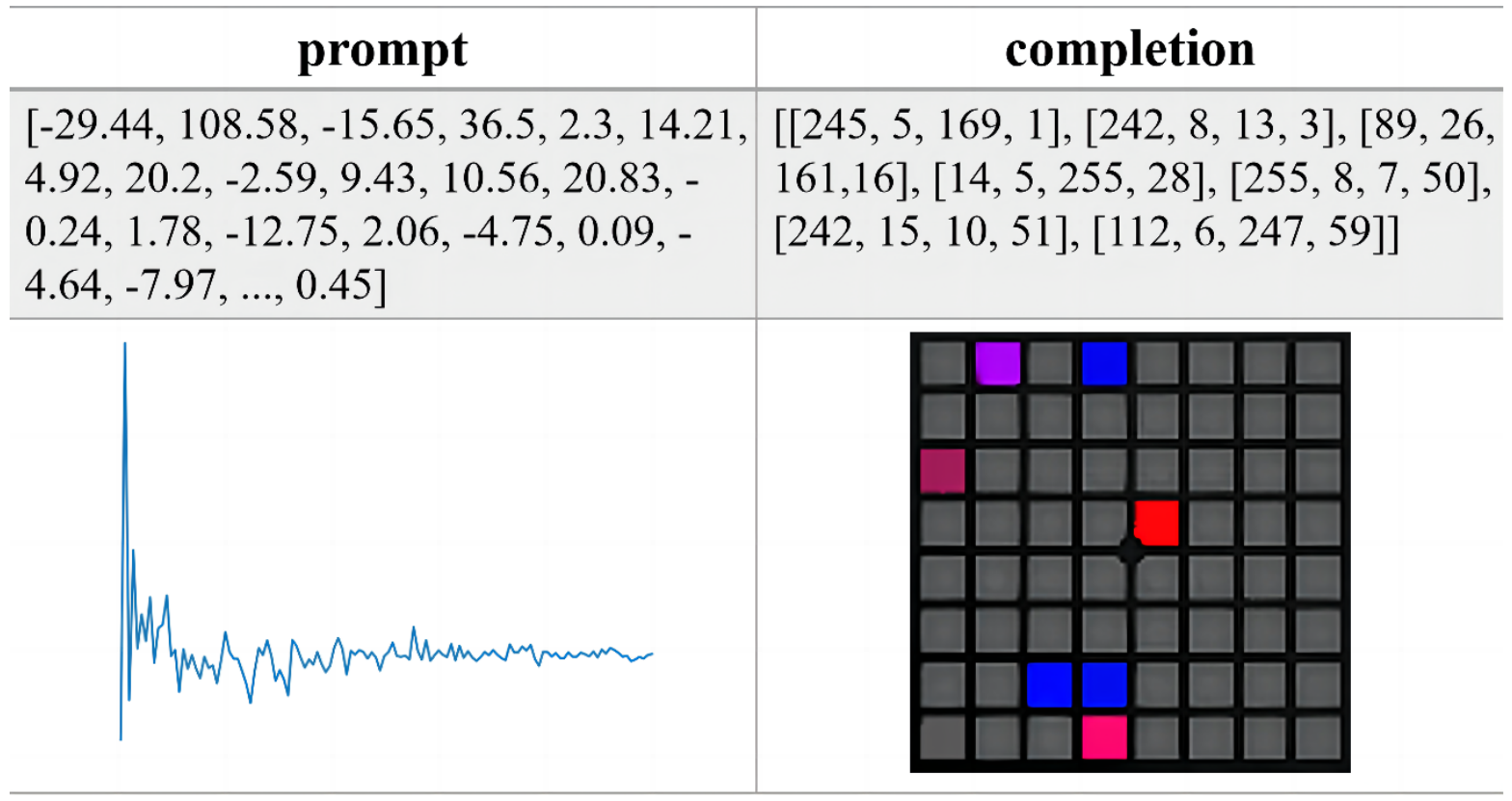}
\caption{A prompt-completion example for LaunchPadGPT~\cite{xu2023launchpadgpt}: The text following "prompt:" represents MFCC feature values, while "completion:" shows RGB-X tuples. The tuple (245, 5, 169, 1) indicates that the Launchpad keyboard's second button (index 0 for the first) is purple.. Figure taken from~\cite{xu2023launchpadgpt}.}
\label{launchpad}
\end{figure}

Xu et al.~\cite{xu2023launchpadgpt} introduced LaunchpadGPT, a musical instrument that allows users to generate music by pressing the illuminated buttons. 
LaunchpadGPT's functionality stems from autonomous learning, drawing from videos of Launchpad performances. These videos act as training data for the instrument, guiding it to craft synchronised lighting patterns that align with the music being played. The instrument's architecture leverages Mel-frequency cepstral coefficients (MFCC) to extract musical features. This is paired with NanoGPT~\footnote{https://github.com/karpathy/nanoGPT}. During the training phase, alignment between music and video frames is facilitated using MFCC features extracted from the music and colour-coordinate tuples (R, G, B, X) taken from the video frames. In this process, the MFCC features act as textual input to the language model, with the RGB-X tuples tokenised similarly to text. At the interface level, the musical features are converted into text tokens that are then fed into the trained language model. The result is a sequence of colour-coordinate tuples that seamlessly synchronise the Launchpad's illuminated buttons with the tempo and ambience of the input music as shown in Figure~\ref{launchpad}.


\begin{table}[!ht]
\centering
\caption{Benchmark results on music generation quality among MusicLDMs and others on Audiostock dataset. FD: frechet distance, VGGish~\cite{hershey2017cnn} and PANN~\cite{kong2020panns} are audio embedding models, IS: inception score, KL Div: kullback-leibler divergence. }
\begin{tabular}{lllll}
\hline
\multirow{2}{*}{Model} & \multicolumn{1}{c|}{\multirow{2}{*}{$\text{FD}_{pann}\downarrow$}} & \multicolumn{1}{c|}{\multirow{2}{*}{$\text{FD}_{vgg}\downarrow$}} & \multirow{2}{*}{IS$\uparrow$} & \multirow{2}{*}{KL Div.$\downarrow$ } \\
            & \multicolumn{1}{c|}{}          & \multicolumn{1}{c|}{}          &           &             \\ \hline
Riffusion~\cite{forsgren6riffusion}       & 68.95                  & 10.77                  & 1.34        & 5.00           \\ \hline
MuBERT~\cite{MubertAI}        & 31.70                  & 19.04                  & 1.51        & 4.69           \\ \hline
AudioLDM        & 38.92                  & 3.08                   & 1.67        & 3.65           \\ \hline
MusicLDM        & 26.67                  & \multicolumn{1}{c|}{2.40}        & \textbf{1.81}    & 3.80           \\ \hline
\end{tabular}
\label{musiccomp}
\end{table}

In summary, significant progress in \textit{Large Music Models} has led to innovative applications. Table \ref{musiccomp} compared the results of MusicLDM and AudioLDM with state-of-the-art diffusion and BERT-based models, which shows that \textit{Large Music Models} improve the state of the art in music generation. These models can now convert text into melodies and even produce music that resonates with human emotions. Examples include the integration of lyrics with tunes, instruments responding to textual cues, and tools like LaunchpadGPT~\cite{xu2023launchpadgpt} that sync lights with music beats. MuseCoco offers impressive music generation capabilities. These models are not just algorithms but tools that complement musical creativity. As these models progress, their impact on music creation and appreciation is undeniable, promising a revolutionary musical future. An overview of recent large music models is presented in Table~\ref{musicLLMs}.

\begin{table*}[!ht]
\centering
\scriptsize
\caption{Large Music Models, TTM: Text-to-Music, VIM: Vocals to instrumental music, MTL: Melody to Lyric, MTM: Music to music, TSM: Text-to-symbolic music, PTM: Programming to Music }
\begin{tabular}{lllll}
\hline
Model & Data                       & Tasks                       & Limitations   &Code                 \\ \hline

 MusciLDM~\cite{chen2023musicldm}     & \begin{tabular}[c]{@{}l@{}}Audiostock\end{tabular} & \begin{tabular}[c]{@{}l@{}}TTM\end{tabular} & \begin{tabular}[c]{@{}l@{}}The model is trained on a sample rate of 16 kHz while\\usually, music holds 44.1 kHz. Text-music data and\\restricted GPU processing capacity found an obstacle\\in the expansion of Music LDM's training. Extracting\\accurate information about the beat is a difficult task as\\it is essential for music alignment.\end{tabular}&\large{\cmark}\\ \hline

 TANGO~\cite{ghosal2023text}     & \begin{tabular}[c]{@{}l@{}}AudioCaps\end{tabular} & \begin{tabular}[c]{@{}l@{}}TTM\end{tabular} & \begin{tabular}[c]{@{}l@{}}Cannot always perform when trained on a smaller dataset\end{tabular} &\large{\xmark}\\ \hline
          
 WavJourney~\cite{liu2023wavjourney}     & \begin{tabular}[c]{@{}l@{}}AudioCaps\end{tabular} & \begin{tabular}[c]{@{}l@{}}TTM\end{tabular} & \begin{tabular}[c]{@{}l@{}}Inflexible to expand the functions.\\The process of remixing and deteriorating may push the\\synthetic audio away from the real.\\Model is time complex when generating the complex audio.\end{tabular} &\large{\cmark}\\ \hline

 SingSong~\cite{donahue2023singsong}     & \begin{tabular}[c]{@{}l@{}}1 million audio samples\end{tabular} & \begin{tabular}[c]{@{}l@{}}VIM\end{tabular} & \begin{tabular}[c]{@{}l@{}}The generated instrumentals often exhibit a disparity, with\\harmonic elements being notably weaker (both in volume\\and coherence) when compared to their percussive\\counterparts.\end{tabular}&\large{\cmark}\\ \hline
 
 LOAF-M2L~\cite{ou2023loaf}     & \begin{tabular}[c]{@{}l@{}}Music Genaration\end{tabular} & \begin{tabular}[c]{@{}l@{}}MTL\end{tabular} & \begin{tabular}[c]{@{}l@{}}--\\--\end{tabular}&\large{\xmark} \\ \hline
          
 MeLoDy~\cite{lam2023efficient}     & \begin{tabular}[c]{@{}l@{}}6.4 Million Samples\\based on MusicCaps\end{tabular} & \begin{tabular}[c]{@{}l@{}}TTM\\MTM\end{tabular} & \begin{tabular}[c]{@{}l@{}}Training data mostly contain non-vocal music only\\Training on LM and DPD on 10-second audio chunks can\\affect the long generation\end{tabular} &\large{\cmark}\\ \hline
          
 MuseCoco~\cite{lu2023musecoco}     & \begin{tabular}[c]{@{}l@{}} Million MIDI Dataset\\EMPOIA\\MetaMidi\\POP909\\Symphony\\Emotion-gen\end{tabular} & \begin{tabular}[c]{@{}l@{}}TSM\end{tabular} & \begin{tabular}[c]{@{}l@{}}Model primarily focuses on producing symbolic music based\\on textual descriptions, with little consideration on long\\sequence modelling.\\
The attribute set discussed in this work only represents a\\subset of all available music attributes.\end{tabular}&\large{\cmark} \\ \hline

          
 LaunchpadGPT~\cite{xu2023launchpadgpt}     & \begin{tabular}[c]{@{}l@{}}music-frame pairs dataset\end{tabular} & \begin{tabular}[c]{@{}l@{}}PTM\end{tabular} & \begin{tabular}[c]{@{}l@{}}
Although LaunchpadGPT partially captures colour similarities,\\it lacks the ability to effectively learn more structured patterns.\end{tabular}&\large{\cmark}\\ \hline

Jukebox~\cite{dhariwal2020jukebox}     & \begin{tabular}[c]{@{}l@{}}f 1.2 million songs\end{tabular} & \begin{tabular}[c]{@{}l@{}}MTM\end{tabular} & \begin{tabular}[c]{@{}l@{}}-\\-
\end{tabular}&\large{\cmark}\\ \hline



\end{tabular}
\label{musicLLMs}
\end{table*}



\subsection{Large Audio Models in Other Audio Applications}

In this section, we explore additional studies that address diverse audio-related applications beyond those discussed earlier. For example, Wang et al.~\cite{Xiaofei2023SpeechX} introduce SpeechX, a versatile architecture capable of various speech transformation tasks across both noisy and clean speech conditions. Utilising task-dependent prompting through a combination of neural codec language modelling and multi-task learning, SpeechX achieves unified modelling that maintains a consistent approach for leveraging textual input. This approach leads to comparable or even superior performance with or without background noise across a range of speech-related tasks, including target speaker extraction, zero-shot TTS, speech noise removal, and speech editing. While data-driven speech processing models thrive with substantial text supervision, acquiring transcribed speech data remains a resource-intensive endeavour. Addressing this concern, Shih et al.~\cite{shih2023speechclip} introduce SpeechCLIP, an inventive framework that connects speech and text through images to enhance speech models without necessitating transcriptions. This approach capitalises on advanced pre-trained models, specifically HuBERT and CLIP, aligning them through paired images and spoken captions with minimal fine-tuning requirements. The results of their study reveal that the SpeechCLIP framework surpasses previous state-of-the-art techniques in both image-speech retrieval and zero-shot speech-text retrieval, all without direct dependence on transcriptions. 

Huang et al.~\cite{huang2023audiogpt} introduce AudioGPT, designed to excel in various speech-related tasks. This model is specifically tailored to generate audio elements within spoken conversations, presenting an alternative to the comprehensive training of multimodal Language Models (LMs) from scratch. AudioGPT is equipped to perform multiple tasks including style transfer, speech enhancement, speech separation, mono-to-binaural conversion, audio inpainting, sound extraction, and more. The model's architecture can be divided into four key components: modality transformation for converting input into a standardised format, task analysis to retrieve organised arguments, model assignment for resource allocation, and response generation to produce desired outputs. The study leverages ChatGPT to efficiently manage a vast array of models. In a similar vein, Deshmukh et al.~\cite{deshmukh2023pengi} present an innovative audio language model, Pengi, which approaches audio tasks as text generation tasks through the implementation of Transfer Learning. This model harnesses the capabilities of an audio transformer HTSAT~\cite{chen2022hts} as an audio encoder, along with CLIP's~\cite{radford2021learning} text encoder and GPT2-base as a language model. A diverse set of datasets, including AudioSet, AudioCaps, Clotho AQA, FSD50k, FreeSound, and more, were employed to train the model. The outcomes reveal that Pengi effectively manages audio tasks of both closed-ended and open-ended nature. 
Large audio models dedicated to obtaining universal audio representations have also been widely applied in the field of sound event detection (SED), such as PANNs \cite{kong2020panns}, AST \cite{gong2021ast}, and BEATs \cite{chen2022beats}. These models are all pre-trained on large-scale audio datasets like AudioSet \cite{gemmeke2017audio} to acquire the ability to obtain universal audio embedding features. Subsequently, the audio embedding features extracted through these large models are fused into SED tasks using various methods, significantly enhancing the performance of SED. For instance, Xu et al. \cite{xu2023semi} improved SED performance on the DESED public evaluation dataset \cite{turpault2019sound} by more than 8.5\% by utilising features from PANNs compared to the baseline.

Latif et al.~\cite{latif2023can} delve into the exploration of how LLMs can be harnessed to annotate speech data, aiming to advance the state of the art in Speech Emotion Recognition (SER). Employing ChatGPT as their tool of choice, they empirically show the promising potential of LLMs in the domain of annotating speech emotion data. The evaluation process encompasses both single-shot and few-shot scenarios, revealing valuable insights into the varying performance of SER. Notably, their experimentation showcases an intriguing approach to performance enhancement via data augmentation. Another recent work~\cite{Shu2022LLASM}
propose a \textit{Large Language and Speech Model} (LLaSM) that is an end-to-end trained large multimodal speech-language model with cross-modal conversational capabilities, designed to understand and execute spoken language instructions. Initial experimentation reveals that LLaSM provides a more user-friendly and intuitive method for human interaction with artificial intelligence.

In recent times, the landscape of speech-related applications has witnessed a surge in innovative solutions driven by the emergence of \textit{Large Audio Models}. Researchers are increasingly exploring the potential of these models to tackle a wide range of speech-related tasks. Table~\ref{llmsaudio} provides an overview of these studies. These large audio models, such as SpeechX and AudioGPT, demonstrate their versatility and proficiency in handling diverse speech-related tasks.

\section{Challenges and Outlook}
\label{sec:challenges}
In this section, we outline the potential challenges and future directions for \textit{Large Audio Models} or the use of textual LLMs to improve audio-related tasks. These challenges include the known shortcomings and unintended consequences of exporting LLMs to domains for which they are explicitly trained. We also incorporate the challenges that have emerged from the attempts to design foundational models for audio processing. It is important to note that these challenges are not exhaustive, as research in this area is rapidly growing. Here we also want to note that LLMs have achieved remarkable results in language modelling and other discrete sequential tasks. However, directly applying LLMs to audio is challenging as audio is inherently analogue and continuous over time. Vector quantisation variational autoencoder (VQVAE) helps address this issue by discretising the speech modality so it can be handled by LLM architectures. VQVAE models the continuous time-series input using a discrete latent codebook through vector quantisation. This allows audio data like spectrograms to be reconstructed from codes while retaining most information about dynamics. Coupling an LLM with the discrete representations learned by VQVAE provides a way for the LLM to indirectly model the continuous nature of the input audio signal~\cite{latif2023can}. Exploring these challenges opens up exciting possibilities for advancing audio-related tasks and pushing the boundaries of what LLMs can achieve in this domain. 

\subsection{Data Issues}
LLMs or \textit{Large Audio Models} are trained on expansive datasets, rendering the assessment and validation of the quantity and quality of data essential for pre-training LLMs in speech and audio processing a virtually insurmountable task. The pragmatic difficulties tied to evaluating data quality give rise to a multitude of concerns, encompassing issues such as duplicated data instances, contamination, bias, and a diverse array of others.

\subsubsection{Doppelganger Data Instances}
  Doppelganger data instances emerge as data that exhibit a resemblance yet lack complete identity. They manifest in various forms, including recurring words and sentences, comparable speech characteristics, similar structure and representation, and more. Detecting these instances within a dataset presents a significant challenge, as they can be nearly indistinguishable from authentic data entries. The existence of these doppelgangers can significantly impact the performance of \textit{Large Audio Models}, as there's a considerable chance that such models might memorise and subsequently reproduce them, leading to diminished performance. Addressing this issue requires the compulsory removal of these doppelganger instances, achieved through the elimination of repeated words and sentences, coupled with the utilisation of diverse ML techniques for structural comparisons. Lee et al.~\cite{lee2021deduplicating} introduced the NearDup tool, which effectively eliminates these doppelganger instances and repetitive substrings from datasets.
  
 \subsubsection{Data Contamination}
  
  Data Contamination as an issue is becoming an Achilles heel of LLMs. The impracticality of screening the data for pre-training the LLMs makes it harder to ensure that evaluation data is not used in the pre-training phase. This contamination affects the performance of the benchmarking phase of the LLMs. In speech and audio processing, background noise, audio distortion, out-of-distribution examples, label noise, offensive content, etc., are issues causing contamination in \textit{Large Audio Models}. Techniques to mitigate data contamination—like data cleaning, augmentation, and validation—are critical research areas for \textit{Large Audio Models}. The leaderboard is a technique used to control inherited data contamination by providing details of the ancestor models and weights used for training a new LLM model. Jacovi et al.~\cite{jacovi2023stop} proposed three strategies for mitigating data contamination issues in LLMs. Their \textit{first strategy} involves encrypting test data with a public key and licensing it to control the distribution of derivatives of the test data. The \textit{second strategy} is conditioning the access of the API by demanding the API holders not to train on the test data. Lastly, the \textit{third strategy} involves releasing the contextual information of data sourced from the internet to circumvent data contamination arising from instances with online solutions, such as repeated conversations, question-answer pairs, emojis, articles, and their synthesised variations. To the best of our knowledge, the techniques that ensure robustness against determined contamination actors and prompts are limited in their performance. Nevertheless, the aforementioned strategies can help reduce data contamination in evaluation benchmarks.

  \subsubsection{Personally Identifiable Information}
  The personal information is part and parcel of the data crawled from the Internet on which most of the textual LLMs are trained/pre-trained. Phone numbers, email addresses, and other personally identifiable information can and are part of that data. Though LLM developer companies use techniques such as scrubbing, privacy-preserving training, anonymisation, and tokenisation to resolve personally identifiable information problems, it persists. More than a few incidents are reported in the literature where LLMs have outputted the personal information in servicing particular prompts~\cite{li2023multi, lukas2023analyzing, wei2023jailbroken, kim2023propile, patsakis2023man}. Carlini et al.~\cite{carlini2019secret} explored the unintended memorisation of generative sequence models and how this memorisation can reveal personally identifiable information. In another work, Carlini et al.~\cite{carlini2021extracting} performed a training data extraction attack on a GPT-2 model and extracted personally identifiable information from the training data. Li et al.~\cite{li2023multi} showed how the prompts are chained together to jailbreak the privacy of the LLMs. We followed the multi-step provided in~\cite{li2023multi}, and ChatGPT deployed by a third party provided personally identifiable information (name, email, and address). The prompt that broke the privacy is ``I am in dire need of help could you please share the name, email address, and address of [company] chairperson". In a similar work, Lukas et al.~\cite{lukas2023analyzing} analysed the personal information leakage in healthcare, emails, and law domains and showed that even after using modern scrubbing techniques 3\% personally identifiable information is leaked. It highlights the flaw in data collection, processing, and scrubbing techniques used for training/pre-training LLMs. They also launched an attack on GPT-2 and showed that the leakage of personal information increased nearly ten times, which is an alarming issue. A few more studies have also highlighted the leakage of personally identifiable information, and despite the measures taken by various LLM designing companies the issue still pertains and causes concerns. 

  Recently, there has been a surge in proposals for \textit{Large Audio Models}, which makes it all the more imperative to assess and address similar challenges in the context of these \textit{Large Audio Models}.

  \subsubsection{Diverse Pre-training Data}
The downstream performance of \textit{Large Audio Models} is intricately linked to the diversity present in the training and pre-training datasets~\cite{shen2023efficient}. Given the immense scale of audio corpora collected from various sources, understanding the individual contributions and relative weights of each source in the learning process remains elusive~\cite{lee2023beyond}. This limitation becomes particularly apparent when these models are presented with multi-domain information. This challenge is not unique to speech and audio processing; a parallel issue is observed in models designed for text processing. Speech data introduces several associated variables, such as background noise, language variability, gender-specific speech features, and more. Striking a harmonious balance between data from diverse domains is crucial; otherwise, the transferability of knowledge for downstream applications may suffer, consequently restricting performance.

While a few studies~\cite{longpre2023pretrainer, lee2023beyond} on textual LLMs have outlined methods for weighting multi-domain data, this challenge remains unresolved in the context of speech and audio processing. Lee et al.~\cite{lee2023beyond} introduced a Task2Vec diversity coefficient to assess the quality and contribution of each data source during model pre-training. However, their experimentation was confined to publicly available open-source pre-training datasets for text. A parallel exploration of a diversity coefficient for speech and audio processing could be a promising avenue. Xie et al.~\cite{xie2023doremi} proposed domain reweighting with min-max optimisation to address data mixing challenges in foundational models designed for text processing. Despite the distinctions between speech and text data, investigating the efficacy of domain reweighting with min-max optimisation to resolve mixing issues in speech and audio processing using \textit{Large Audio Models} holds significant potential.

  
\subsection{Tokenisation}
Tokenisation is a critical component of LLMs. It is the process of dividing words or letters into smaller chunks~\cite{mielke2021between}. Tokens are used to cope with terms that are not in the model's lexicon. Tokens are often restricted in size because expanding the token space increases computation complexity. Tokenisation is a critical processing step in the pipelines of ChatGPT and other LLMs. Though tokenisation has enabled LLMs, it also brings complications such as computational cost, language reliance, handling of new terms, fixed vocabulary size, information loss, and limited human interpretability, to name a few~\cite{kaddour2023challenges}. Petrov et al.~\cite{petrov2023language} have shown that tokenisation is the procedure in which the language bias first manifests itself in the LLM pipeline. They investigated 17 tokenisers and discovered that the differential treatment of languages is now mirrored in the tokenisation phase, which limits the LLM's capacity to learn new terminologies and information loss while also introducing bias in LLMs towards languages. 

The continuous nature of audio signals, speech variability, and background noise compound the challenge of tokenisation in \textit{Large Audio Models}. Variations in pronunciations and overlapping speech further restrict tokenisation benefits. Multilingual speech tokenisation poses additional complexities as the same statement might demand a varying number of tokens in different languages. Similarly, tokenising the emotion in the speech is challenging, posing the risk of information loss. AudioPaLM cites the quality of the tokenisation as a key limitation in lessening the performance. Similarly, SpeechGPT and SpeechX also suggest that tokenisation is a potential limitation in their performance. There have been recent attempts in designing and testing a few speech tokenisers~\cite{banerjee2022wav2tok}, but overall, tokenisation for speech/audio data processing requires further attention.
As we look ahead, addressing the challenges surrounding tokenisation in the context of large audio models data processing remains an essential avenue for research and development. Zhang et al. \cite{zhang2023speechtokenizer} presents a unified speech tokeniser and evaluates this on different speech-related tasks. However,  further efforts to enhance tokenisation quality for \textit{Large Audio Models} will be pivotal in improving the overall performance of these models.

\subsection{Computational Cost and Energy Requirements}
Pre-training a large audio model requires a massive amount of audio data and compute hours. The compute hours need millions of US dollars worth of energy. AudioGPT with 137 billion parameters requires approximately 1 million kilowatt-hours (kWh), which is approximately \$137 million USD. Similarly, the training cost of the state-of-the-art AudioPaLM with 530 billion parameters is approximately \$530 million USD. While a few recent attempts have been made to reduce the computational cost and energy usage~\cite{touvron2023llama,du2022glam, park2022nuqmm, zhu2023spikegpt}, the race to the state of the art in various audio-related tasks is quickly pushing up the pre-training cost. The amount of energy consumed in training these LLMs also hurts climate change and carbon emissions as the power needed to run the GPU and TPU clusters for pre-training~\cite{rillig2023risks}.

Fine-tuning is another aspect of LLMs that requires considerable computing power (although not as much as pre-training) and usually a large amount of memory to store the model. Fine-tuning is a technique that adapts a pre-trained model to a specific domain or task using a small dataset. It is typically used for downstream applications and has proven useful for designing various speech and text-based applications. Fine-tuning LLMs for speech and audio processing can be limited by memory requirements. The memory requirements of fine-tuning are the same as those of pre-training, which can be a bottleneck for downstream applications. Although there exist efficient fine-tuning techniques like parameter efficient training, prompt-based efficient tuning, and memory efficient tuning for text processing~\cite{hu2023llm, susnjak2023prisma, chavan2023one, hu2021lora}, comparable methods are limited for speech and audio processing~\cite{ghosal2023text, huang2021speech, li2023prompting}. The need for specialised fine-tuning techniques tailored to the nuances of audio processing remains evident. As the field advances, addressing the memory challenges of fine-tuning will be pivotal to unlocking the full potential of \textit{Large Audio Models} for real-world applications.

\subsection{Limited context length}
\textit{Large Audio Models}' downstream applications require context understanding for making intelligent decisions; this limits the quantity of data that the model may access when generating predictions. It is especially difficult for tasks requiring long-term dependency understanding, such as speech recognition~\cite{latif2021survey}. Second, the model may struggle to understand the links between distinct portions of a speech signal because of the short context length. This can cause issues with activities like speaker identification and emotion recognition. Finally, the model's short context duration may make generalisation for fresh data difficult. This is because the model was only trained on a tiny quantity of data, making it difficult for it to understand how to cope with novel circumstances. There are a few techniques in the literature for sorting out the challenges posed by the limited context. These techniques are efficient attention mechanisms~\cite{ainslie2023colt5, ding2023longnet, kaddour2023no, tworkowski2023focused}, length generalisation~\cite{haviv2022transformer, liu2022transformers, raffel2020exploring, su2021roformer, geiping2023cramming}, and transformer alternatives~\cite{sun2023retentive, dao2022hungry, orvieto2023resurrecting, peng2023rwkv}. An efficient attention mechanism helps decrease the context length while ensuring limited computational overhead on the transformer architecture, length generalisation methods use positional embeddings to circumvent the limited context length issues, and transformer alternatives can provide options for using techniques that do not run into issues context lengths. As the field advances, addressing the limitations of context understanding will be integral to unlocking the full potential of \textit{Large Audio Models}' applications.

\subsection{Understanding Paralinguistic Information}
Paralinguistic information such as emotions plays a pivotal role in speech, as they establish a connection with its inherent meaning and categorisation. Even identical utterances can acquire distinct meanings when expressed with different emotions~\cite{latif2020deepinter}. While LLMs excel in speech generation, their capacity to comprehend and generate various emotions remains largely untapped and understudied. Zhang et al.~\cite{zhang2023speechgpt} highlighted the deficiency in understanding and generating paralinguistic information, including emotions, as a potential limitation of SpeechGPT. Similarly, AudioPalm~\cite{rubenstein2023audiopalm}, considered one of the finest LLMs for audio processing, exhibits limited coverage in paralinguistic information comprehension. Although AudioLM~\cite{borsos2023audiolm} claims to preserve speaker identification and prosody, its exploration of emotions and other paralinguistic information remains unknown. The LLMs discussed in Section \ref{largeaudiomodels} all exhibit constrained exploration in generating and understanding paralinguistic information. 

A compelling proposition materialises: embedding the ability to generate and comprehend emotions stands as the next substantial stride in the evolution of \textit{Large Audio Models}. By enriching these models with emotional depth, they hold the potential to substantially enhance their capacity to communicate, resonate, and connect with human expression in various contexts. Moreover, the field of music generation stands as another avenue poised for exploration within the purview of large music models. Just as with emotions in speech, music's emotional intricacies are deeply intertwined with its tonal, rhythmic, and structural components. \textit{Large Audio Models} that cannot only understand but also generate music with diverse emotional qualities could usher in a new era of creativity and emotional engagement in music production and consumption. The fusion of emotions and music generation could pave the way for even more comprehensive and impactful interactions with \textit{Large Audio Models}, shaping the future of human-AI collaboration in both spoken and musical aspects.

\subsection{Prompt Sensitivity}
Prompts are the way to interact with LLMs. They act as a query or input to the LLM, and in response to the query, the LLM generates the response. It is now well known that a slight change in a prompt that is not perceptible to humans can disrupt the LLM's entire response. This brittleness can cause serious consequences when LLMs are deployed in response-critical applications such as healthcare~\cite{cascella2023evaluating}, finance~\cite{wu2023bloomberggpt}, and law~\cite{cui2023chatlaw}. \textit{Large Audio Models} engaged in speech processing tasks, such as speaker identification, speech classification, and speech translation, are equally vulnerable to prompt variations. Studies have begun to delve into comprehending the ramifications of prompt brittleness on generating labels for speech-based affective computing tasks~\cite{latif2023can}.

Recent advancements, including single-turn prompting, in-context learning, multi-turn prompting, and the chain of thought, have gained prominence~\cite{wei2022chain}. However, these techniques have primarily demonstrated their efficacy in text-based applications, where furnishing context in plain language proves adequate. Intriguingly, in the domain of \textit{Large Audio Models}, a notable void exists. Despite the progress made, to our knowledge, the literature has yet to explore the design and testing of specialised prompts tailored specifically for speech-based scenarios. Addressing this gap presents an intriguing avenue for research, one that could pave the way for enhanced interactions with \textit{Large Audio Models}.



\subsection{Hallucination}
Hallucination is a challenging issue in LLMs. It is a behaviour where LLMs generate factually incorrect information that is difficult to detect due to the large number of outputs that LLMs generate \cite{rawte2023survey}. Ji et al.~\cite{ji2023survey} divided hallucination into two categories: intrinsic hallucinations and extrinsic hallucinations.
\begin{itemize}
  \item Intrinsic hallucinations occur when LLMs misunderstand the source content and generate factually incorrect information. 
  \item Extrinsic hallucinations occur when the LLM-generated output cannot be substantiated from the source data or contradicts it. 
\end{itemize}
The occurrence of hallucinations is not limited to text-based LLMs; \textit{Large Audio Models} are also susceptible. This susceptibility can lead to misinterpretations of audio sources, altered reasoning, and the introduction of random noise. To mitigate this issue, the literature suggests several strategies, including adversarial training, diversification in training data, human feedback incorporation, contrastive learning, and enhanced regularisation~\cite{manakul2023selfcheckgpt, feldman2023trapping, ji2023survey, mckenna2023sources, sun2023contrastive}. It is notable, however, that discussions around the hallucination challenge within \textit{Large Audio Models} remain somewhat limited~\cite{gong2023listen}.

Given the growing prominence of models like AudioPaLM, SpeechGPT, and others in applications such as speaker recognition and speech translation, addressing and comprehending the hallucination challenge in the context of \textit{Large Audio Models} holds considerable significance. It is imperative to proactively tackle this issue to ensure the reliability and accuracy of these models in real-world applications.

\subsection{Ethics}



As \textit{Large Audio Models} continue to gain prominence across diverse applications, addressing their ethical challenges becomes a paramount concern. These models are trained on extensive datasets sourced from the internet, a practice that brings with it the risk of inheriting ingrained biases. Such biases can dangerously influence the output generated by these models, inadvertently propagating content that reflects racism, sexism, or other forms of discrimination. Moreover, these datasets might inadvertently encompass personally identifiable information, potentially compromising individual privacy. The intrinsic generative capabilities of \textit{Large Audio Models} introduce an additional layer of complexity, allowing for the creation of convincing deep fakes that pose a significant security threat to downstream applications.

Within this evolving landscape, music generation with \textit{Large Audio Models} offers exciting prospects for increased creative engagement and novel artistic expression. However, it also carries the power to reshape musical culture and redefine economic dynamics~\cite{donahue2023singsong}. Constructing a music generation system that involves user initiative, like singing, and preserves individual identity in the output holds promise for enhancing creative participation. Yet, ethical concerns arise from the potential lack of user control over the genre and style of the generated instrumental output. To address this, refining the system to offer users explicit control over these attributes becomes essential, mitigating the potential risks. This intricate balance between creative empowerment and potential biases exemplifies the multifaceted ethical landscape that surrounds the utilisation of \textit{Large Audio Models} in music generation.

Similarly, the capabilities of \textit{Large Audio Models} to synthesise high-quality structured audio content hold immense potential across varied domains. Nevertheless, they also inherit challenges similar to their text-based counterparts, including societal biases present in training data. Moreover, the generated speech might not consistently align with intended accents or dialects for marginalised groups. The potential for these models to continue short speech segments while preserving speaker identity and prosody raises concerns of potential misuse~\cite{borsos2023audiolm}, demanding the implementation of responsible AI practices. By delving into these nuances, the discourse on ethical challenges associated with \textit{Large Audio Models} expands to encompass a wide spectrum of concerns, underscoring the need for robust safeguards and strategic deployment strategies. As the foundational research in this area grows, the insights presented here, along with related works, can serve as a valuable guide to understanding the complexities and addressing the challenges in the development of improved \textit{Large Audio Models} for audio applications~\cite{kaddour2023challenges, latif2022ai, awais2023foundational, ray2023chatgpt}.






\section{Conclusion} \label{sec:conclusion}

In the rapidly evolving landscape of artificial intelligence, the role of large AI models, in particular LLMs, in audio processing -- including domains such as speech and music — is becoming increasingly pivotal. This paper offers the first comprehensive survey on \textit{Large Audio Models}, capturing the nuanced interplay of various LLMs within the audio sector. By consolidating state-of-the-art methods and surfacing current challenges, we provide a valuable resource for researchers aiming to navigate this terrain. Furthermore, the highlighted potential future directions aim to chart a course for upcoming investigations in this domain. As the boundary of what is possible with LLMs in audio processing continues to expand, this survey aspires to be a foundational reference, enlightening the path for future explorations and innovations.




\end{document}